\documentclass[twocolumn]{aa} 

\usepackage{graphicx}
\usepackage{txfonts}
\usepackage[normalem]{ulem}
\usepackage{color}
\usepackage{esvect}

\definecolor{purple}{rgb}{0.5,0,0.87}

\newcommand\modified[1]{} 

\usepackage[colorlinks,allcolors=blue]{hyperref}

\titlerunning{Global potential of low mass galaxies from observed surface brightness}
\authorrunning{S\'anchez~Almeida et al.}

\begin{document} 
   \title{Application of the Eddington inversion method to constrain the dark matter halo of galaxies  using only observed surface brightness profiles}
   \author{
     Jorge S\'anchez Almeida\inst{1,2}
     \and
     Angel R. Plastino\inst{3}
     \and
     Ignacio Trujillo\inst{1,2}
          }
          \institute{Instituto de Astrof\'\i sica de Canarias, c/ Vía Láctea s/n, E-38205, La Laguna, Tenerife,  Spain
\and
Departamento de Astrof\'\i sica, Universidad de La Laguna, E-38203, La Laguna, Tenerife, Spain
\and
CeBio y Departamento de Ciencias B\'asicas, Universidad Nacional del Noroeste de la Prov. de Buenos Aires, UNNOBA, CONICET, Roque Saenz Peña 456, Junin, Argentina
      }

   \date{Received \today; accepted March, 14, 1592}

\abstract
{The halos of low-mass galaxies may allow us to constrain the nature of dark matter (DM), but the kinematic measurements to diagnose the required properties are technically extremely challenging. However, the photometry of these systems is doable.}
{Using only stellar photometry, constrain key properties of the DM haloes in low-mass galaxies.}
{Unphysical pairs of DM gravitational potentials and starlight distributions can be identified if the pair requires a distribution function $f$ that is negative somewhere in the phase space. We use the classical Eddington inversion method (EIM) to compute $f$ for a battery of DM gravitational potentials and $\sim 100$ observed low-mass galaxies with $M_\star$ between   $10^6$ and  $10^8\,{\rm M_\odot}$. The battery includes NFW potentials (expected from cold DM) and potentials stemming from cored mass distributions (expected in many alternatives to cold DM).  The method \modified{assumes spherical symmetry and isotropic velocity distribution and} requires fitting the observed profiles with analytic functions, for which we use polytropes (with zero inner slope, a.k.a. core) and profiles with variable inner and outer slopes. \modified{The validity of all these assumptions is analyzed.}
}
{In general, the polytropes fit well the observed starlight profiles. If they were the correct fits (which could be the case) then all galaxies are inconsistent with NFW-like potentials. Alternatively, when the inner slope is allowed to vary for fitting, between 40\,\% and 70\,\% of the galaxies are consistent with cores in the stellar mass distribution and thus inconsistent with NFW-like potentials.}
{Even though the stellar mass of the observed galaxies is still not low enough to constrain the nature of DM, this work shows the practical feasibility of the EIM technique to infer DM properties only from photometry.} 
%
%
   \maketitle

%
\section{Introduction}\label{sec:intro}
The classical Eddington Inversion Method \citep[EIM; ][]{1916MNRAS..76..572E,2006ApJ...642..752A,2008gady.book.....B,2018JCAP...09..040L,2021isd..book.....C}  can be used to constrain the shape of the dark matter (DM) halo of a galaxy using only the observed stellar mass distribution. It provides the distribution function (DF) in the 6D phase space to be followed by a given mass density profile immerse in a given gravitational potential. The minimal requirement for any feasible DF is to be non-negative everywhere in the phase space, and this seemingly simple constrains is enough to discard as unphysical many combinations of density and potential leading to a DF $< 0$. Recently, \citet{2023ApJ...954..153S}  use the  EIM in the regime of low mass galaxies, where the contribution of the baryons to the overall potential can be neglected and the potential is set only by the DM. Among others, they showed that cored stellar mass distributions (i.e., with constant density when the distance to the center goes to zero) are inconsistent with a Navarro, Frenk, and White potential, expected in DM dominated systems with collision-less cold DM \citep[the so-called NFW profiles after][]{1997ApJ...490..493N}. Strictly speaking, this result holds for spherically symmetric particle systems with an isotropic velocity distribution, however, extensions that relax these assumptions and yield similar constraints exist (a detailed account is given in Sect.~\ref{sec:eddington_summary}).

The technique has great potential since, using only photometry, it may be able to constrain the shape of the DM halos in ultra-low mass galaxies. In this mass regime the baryon feedback is unable to turn cuspy DM halos into cored halos. Thus, if the DM haloes of these galaxies happen to have a core, it will indicate a departure of the DM from being collision-less, reflecting the much sought-after presently-unknown true nature of DM \citep[whether it is fuzzy, self-interacting, warm, or other alternatives; e.g.,][]{1994PhRvL..72...17D,2000PhRvL..85.1158H,2000PhRvL..84.3760S, 2022arXiv220307354B,2023arXiv230603903C}. The limit defining ultra-low mass galaxies, i.e.,  the largest stellar mass unable to modify the DM profile, is model dependent \citep[e.g.,][]{2016MNRAS.459.2573R}, however, it roughly corresponds to a stellar mass of $M_\star < 10^{6}\,{\rm M}_{\odot}$ or a DM halo mass of $M_h < 10^{10}\,{\rm M_\odot}$  \cite[e.g.,][]{2014MNRAS.437..415D,2015MNRAS.454.2981C,2020ApJ...904...45H,2021MNRAS.502.4262J,2023MNRAS.519.4384E}. Traditionally, halo shapes are deduced from kinematical measurements, which require high spectral resolution spectroscopy and which, keeping in mind the need for large statistics to derive firm  conclusions on the nature of DM, are virtually imposible in this ultra-low mass regime. However, the broad-band photometry needed to infer their stellar mass profiles starts to be doable \citep[e.g.,][]{2021A&A...654A..40T} and will become routinely simple in the near future with instruments like the Rubin Observatory \citep{2019ApJ...873..111I}. Here is where the EIM comes into play, constraining the properties of the halo mass distribution from the stellar light alone.   

The present work represents the first application to real galaxies the EIM-based technique devised by \citet{2023ApJ...954..153S}. We put forward diagnostic diagrams and then discuss the kind of constraints to be expected in practice. The target galaxies were collected by  \citet{2021ApJ...922..267C, 2022ApJ...933...47C} who select them as dwarf satellites of Milky Way-like galaxies. This sample is not ideal for ascertaining the nature of DM since (1) their masses exceed the $10^6\,{\rm M}_\odot$ limit and (2) the fact that they are satellites complicates the interpretation of the DM halo shape in terms of the DM nature alone. The evolution of satellites may be  influenced by the central galaxy through tidal forces and gas pressure gradients \citep[e.g.,][]{2016MNRAS.459.3998L,2017ApJ...835..159S}. However, this galaxy set is a good testbed since it comprises a large number of objects ($\gtrsim$\,100) with density profiles comparable in lack of symmetry and noise with those to be expected soon for the appropriate galaxies.  

The paper is structured as follows: Sect.~\ref{sec:method} summarizes the technique to be employed, with Sect.~\ref{sec:eddington_summary} giving an overview of the EIM,
including an extension that goes beyond the spherical symmetry assumption (Appendix~\ref{app:extension}).
The actual algorithm is described in Sect.~\ref{sec:procedure} and tested in Appendix~\ref{app:gctests}. The  data are presented in Sect.~\ref{sec:observations}, with the new diagnostic diagrams spelled out in Sect.~\ref{sec:diagrams}.
These diagrams depend on the functions used for fitting the observed profiles. For this task, we firstly use projected polytropes 
in  Sect.~\ref{sec:pp_results}. Polytropes have cores, an assumption relaxed in Sect.~\ref{sec:intermediate} (free inner slope fits) and Sect.~\ref{sec:full} (free inner and outer slope fits). The conclusions arising from the results are analyzed in Sect.~\ref{sec:conclusions} including the impact of assumptions like spherical symmetry and isotropic velocity distributions made when we applied the EIM.

%
\section{The Eddington inversion method (EIM)}\label{sec:method}

For the sake of comprehensiveness, and to set up the scene, Sect.~\ref{sec:eddington_summary} gives a brief account of the EIM and the implications for low mass galaxies found by \citet{2023ApJ...954..153S}. Then the practical application of the method carried out in the present paper is explained  in Sect.~\ref{sec:procedure}.

\subsection{Contextual briefing of the EIM}\label{sec:eddington_summary}

For spherically symmetric systems of particles with isotropic velocity distribution, the phase-space DF $f(\epsilon)$ depends only on the particle energy $\epsilon$. The space density $\rho(r)$ is an integral over all velocities $v$ at a fixed distance $r$. If $\Psi(r) = \Phi_0 - \Phi(r)$ is the relative potential energy, then the relative energy (per unit mass) is $\epsilon = \Psi - \frac{1}{2} v^2$, so that the density becomes \citep[e.g.,][Sect.~4.3]{2008gady.book.....B},
\begin{equation}
  \rho(r) = 4 \pi \sqrt{2} \,\int_0^{\Psi(r)} \, f(\epsilon) \sqrt{\Psi(r) - \epsilon} \, d\epsilon,
  \label{eq:leading}
\end{equation}
with
$\Phi(r)$ the gravitational potential  and $\Phi_0$ its value at the edge of the system.
For realistic systems, the relative potential $\Psi$ is a monotonically decreasing function
of the distance from the center $r$. Consequently, $\rho$ can be regarded as a function of $\Psi$. Differentiating $\rho$
with respect to $\Psi$ yields,
\begin{equation} 
\frac{d \rho}{d \Psi} \, = \, 2\pi \sqrt{2}  \, \int_0^{\Psi} \, \frac{f(\epsilon)}{\sqrt{\Psi - \epsilon}}
\, d\epsilon.
\label{derodepsi}
\end{equation}
Inverting this Abel integral and after some mathematical manipulations, one finds the equation for
$f(\epsilon)$ in terms of the spatial density,
\begin{equation} 
f(\epsilon) = \frac{1}{\sqrt{2} \pi^2} \,
\int_0^{\epsilon} \, \frac{d^3\rho}{d\Psi^3} \, \sqrt{\epsilon - \Psi} \, d\Psi,
\label{eq:ff1}
\end{equation}
which provides the renowned EIM equation. It gives the DF $f(\epsilon)$ consistent with  $\rho$ and $\Phi$.

However, given an arbitrary pair of $\rho$ and $\Phi$ (or $\rho$ and $\Psi$), it is not  guaranteed that the DF yielded by Eq.~(\ref{eq:ff1}) is positive everywhere in the phase space. If $f < 0$ somewhere it means that this particular combination of $\rho$ and $\Phi$ cannot exist in practice. \citet{2023ApJ...954..153S} analyze a sizable set of pairs $\rho$ -- $\Phi$ showing a number of inconsistencies that may be important for real galaxies. The simplest one, yet very powerful,  directly follows from Eq.~(\ref{derodepsi}).  If $d\rho/ d\Psi =0$ then the integral in the right hand side of Eq.~(\ref{derodepsi}) has to be zero so that $f(\epsilon) < 0$ somewhere within the interval $0\le \epsilon\le \Psi$.  This is precisely what happens with any cored density profile immersed in a NFW potential. Since the only variable is $r$, 
 \begin{equation}
   \frac{d\rho}{d\Psi} \, = \, \frac{d\rho/dr}{d\Psi/dr}.
   \label{eq:masterme}
 \end{equation}
A cored mass density is defined to have
 \begin{equation}
  \lim_{r\to 0} \frac{d\rho}{dr} = 0,
   \label{eq:strict}
 \end{equation}
which is inconsistent with a NFW background potential, which has
\begin{equation}
  \lim_{r\to 0}\frac{d\Psi}{dr} \ne 0.
  \label{eq:non-zero}
\end{equation}

The above conclusion holds for spherically symmetric systems with isotropic velocity distribution, however, this restriction can be relaxed to accomodate more general systems. Surpassing the simplest assumption requires treating particular cases individually, each one with its own peculiarities, therefore, only a handful of them have been analyzed so far, probably representing the tip of the iceberg. In order to discuss these individual cases, it is necessary to define the velocity field anisotropy parameter $\beta$,
\begin{equation}
\beta \, = \, 1 \, - \frac{\sigma_{vt}^2 }{2\sigma_{vr}^2},
\label{eq:ani-param}
\end{equation}
where $\sigma_{vr}$  and $\sigma_{vt}$  are the radial and tangential velocity dispersion, respectively. The isotropic systems leading to the EIM Eq.~(\ref{eq:ff1}) have $\sigma_{vt}^2 =2\sigma_{vr}^2$ and so $\beta=0$. Systems with radially biased orbits have $\beta > 0$ whereas systems with tangentially biased orbits have   $\beta < 0$.  Systems with circular orbits are a limiting case of the latter having $\sigma_{vr}=0$ and therefore having $\beta=-\infty$. The case where
 \begin{equation} 
   \beta(r) = \frac{r^2}{r^2 + r_b^2},
   \label{eq:ombeta}
 \end{equation}
 is called Osipkov-Merritt model and introduces a new characteristic length scale $r_b$. It is tractable analytically and approximately describes the global trend expected in low mass galaxies, with $\beta\sim 0$ in the center and then increasing outwards ($\beta >0$).

The constraints on the potential imposed by the stellar mass distribution are best described in terms of the 
 inner logarithmic slope of the stellar density profile $c$,
\begin{equation}
  \lim_{r\to 0} \frac{d\log\rho}{d\log r} = -c,
  \label{eq:innerlogs}
\end{equation}
and the inner slope of the total mass density profile $\rho_p$ defining the gravitational potential,
\begin{equation}
\lim_{r\to 0} \frac{d\log\rho_p}{d\log r} = -c_p. 
\end{equation}
The univocal association between $\Psi$ and $\rho_p$ is granted through the Poisson's equation which, for spherically symmetric systems, yields \citep[e.g.,][]{2013MNRAS.428.2805A},
\begin{equation}
  \Psi =4\pi G\,\left[\frac{1}{r}\,\int_0^r\,t^2\,\rho_p(t)\,dt+\int_r^{\infty}\,t\,\rho_p(t)\,dt\right].
  \label{eq:pot_general}
\end{equation}
In terms of $c$ and $c_p$, a cored stellar density profile has $c=0$ whereas a NFW potential has $c_p=1$. 
From now on, the term {\em soft-core} is used to denote those profiles where the inner slope is not exactly zero but close to it ($c$ or $c_p \gtrsim 0$).
Equipped with these definitions, the main known constraints imposed by the EIM are \citep{2023ApJ...954..153S}:
\begin{itemize}
\item Stellar cores  ($c=0$) and NFW potentials are incompatible,  provided the velocity distribution is isotropic ($\beta=0$). This was already proven above.
\item Stellar cores are also incompatible with potentials stemming from a total density with a quasi-core ($c_p >0$) for $\beta =0$.
\item As expected for physical consistency, stellar cores ($c=0$) and potentials resulting from cored density profiles ($c_p=0$) are consistent in isotropic ($\beta=0$) and radially biased systems ($\beta >0$).  
\item Stellar cores and NFW potentials are also incompatible in systems with anisotropic velocities  following the Osipkov-Merritt model.
\item  Stellar cores and NFW potentials are incompatible in systems with radially biased orbits (${\rm constant~}\beta > 0$). Actually, a stellar core is incompatible with any potential without a core in systems with constant radially biased orbits ($\beta > 0$).
\item Circular orbits ($\beta=-\infty$) can accommodate any combination of baryon density and potential, including a cored stellar density in a NFW potential. However, this configuration is very artificial from a physical standpoint.
\item
A convex linear combination of two DFs that are both consistent with a given potential constitutes a new DF which is also consistent with the same potential.
  Thus, one may think that adding a positive DF for circular orbits may compensate a negative DF for isotropic orbits to yield a positive physically sensible DF. However, this is not the case.  Independently of the relative weight, the superposition of an unphysical DF for isotropic velocities ($\beta=0$) with a physically realizable DF for circular orbits ($\beta =-\infty$) always yields unphysical DFs.
\item Soft-cores are inconsistent with NFW potentials when  $c \lesssim 0.1$ while they are consistent when  $c \gtrsim 0.1$.
\item The inner slope of a soft stellar core and the radial anisotropy are related so that $c > 2\beta$ for the system to be consistent.  In other words, systems with radially biased orbits are inconsistent with soft stellar cores.  
\item Positive inner slope in the stellar distribution ($c < 0$), where the density grows outwards,  is discarded in every way for isotropic velocity. 
\item For stellar densities and potentials with the same shape (whether cored or not), the stellar density distribution cannot be broader than twice the width of the density characterizing the potential. This result refers to isotropic velocities and may be used in real galaxies to set a lower limit to the size of the DM halo from the size of the observed starlight.
\item The above results do not depend on a global  factor scaling the stellar density profile, therefore, they do not depend on the (unknown) ratio between the stellar mass and the total mass of the system. Moreover, it implies that  light profiles or star number counts could replace mass density in the EIM. 
\end{itemize}
To the above list of known constraints, here we add another one relaxing the  l symmetry assumption. In    Appendix~\ref{app:extension}, we prove that the incompatibility between NFW potentials and stellar cores also holds for axi-symmetric systems of arbitrary axial ratio. It is particularly relevant because this fact shows how the incompatibility goes beyond the spherical symmetry assumption, being more fundamental than, and not attached to, this particular hypothesis.

%
%
\subsection{Procedure followed in this paper}\label{sec:procedure}

As summarized in Sect.~\ref{sec:eddington_summary}, the EIM gives the DF $f$ needed for the stellar density profile $\rho$  to reside in a gravitational potential $\Phi$. If two arbitrary $\rho$ and $\Phi$ yield  $f < 0$ somewhere in the phase space, this particular combination is physically inconsistent and can be discarded. Here we apply the technique to constrain the properties of the haloes in a number of observed galaxies. The procedure works as follows:
\begin{enumerate}
\item  Measure $\Sigma_o$ in a real galaxy, i.e., the projection along the line-of-sight (LOS) of the density $\rho_o$. 
\item Fit $\Sigma_o$ with synthetic $\Sigma$ to estimate $\rho_o$, i.e., the $\rho$  whose $\Sigma$ best-fits $\Sigma_o$.  For spherically symmetric systems, the relation between $\rho$ and $\Sigma$ is univocal \citep[e.g.,][]{2008gady.book.....B},
  \begin{equation}
    \Sigma(R) = 2\,\int_R^\infty \frac{r\,\rho(r)}{\sqrt{R^2-r^2}}\,dr,
    \label{eq:projection}
  \end{equation}
  with $R$ the projected distance to center of the system.
\item Using the EIM (Eq.~[\ref{eq:ff1}]), compute the $f_{oi}$ corresponding to the $\rho$ best fitting $\rho_o$ and a battery of $\Phi_i$, with $i=1, 2, \dots$
\item Discards as unphysical, all the  $\Phi_i$ that yield $f_{oi} <0$ somewhere in the phase space.
\end{enumerate}

The above algorithm can be applied to individual galaxies as well as to sets of them ($o=1, 2, \dots$). For example, one can compute the fraction of times that a fixed $\Phi_i$ is unphysical considering all the galaxies in a particular set. This is the approach followed in our work, which requires the computation of $f_{oi}$ for all galaxies in the set ($\forall~ o$) and for all potentials ($\forall~i$). Then we use as diagnostic for the reliability of the potential $\Phi_i$ the fraction of times that it is consistent, i.e., the fraction of times where  $f_{oi} \geq 0$ everywhere. Practical examples of the procedure are given in Sect.~\ref{sec:diagrams}.

  This scheme has been tested for consistency using the set of Globular Clusters (GCs) collected by \citet[][]{2013ApJ...774..151M}. In general, GCs are known to have little or no DM \citep[e.g.,][]{2012ApJ...755..156S,2013MNRAS.428.3648I}, therefore,  GC  can be regarded as self-gravitating systems so that the gravitational potential is set directly by the observed stellar distribution. Moreover, the stellar surface density of GCs usually show cores.  Thus, GCs are excellent testbeds for our EIM-based approach since it provides constraints on the properties of the potential which can be compared directly with the true values obtained from the surface density profile. The tests are detailed in Appendix~\ref{app:gctests}, and they lead to the conclusion that our scheme gives consistent results with the gravitational potential of the GCs inferred from their stellar distribution.

%
%
\begin{figure}
\centering 
\includegraphics[width=1.0\linewidth]{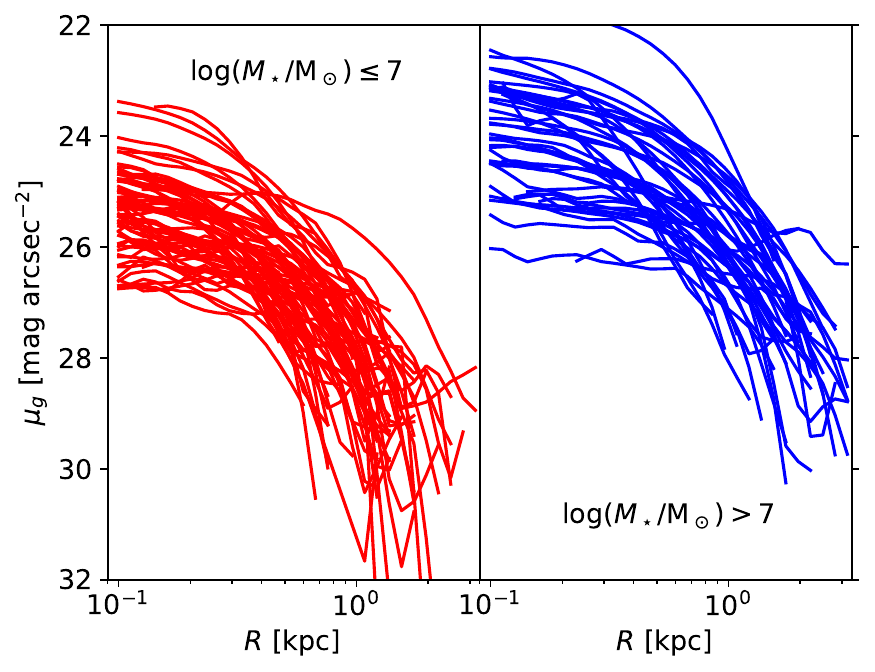}
\caption{
  Surface brightness profiles in the $g$ band for the early-type galaxies in \citet{2021ApJ...922..267C, 2022ApJ...933...47C}. The left panel shows stellar masses between  $10^6$ and $10^7\,{\rm M_\odot}$ whereas the right panel corresponds to  masses between $10^7$ and $10^8\,{\rm M_\odot}$.  Note how the profiles tend to present a central plateau or core. The color code is maintained in Figs.~\ref{fig:eddington1_plot_new_b}, \ref{fig:eddington2_plot_new_b}, and \ref{fig:eddington3_plot_new_b}.
}
\label{fig:raw_prof}
\end{figure}
%
%
\begin{figure*}
  \centering
\includegraphics[width=0.4\linewidth]{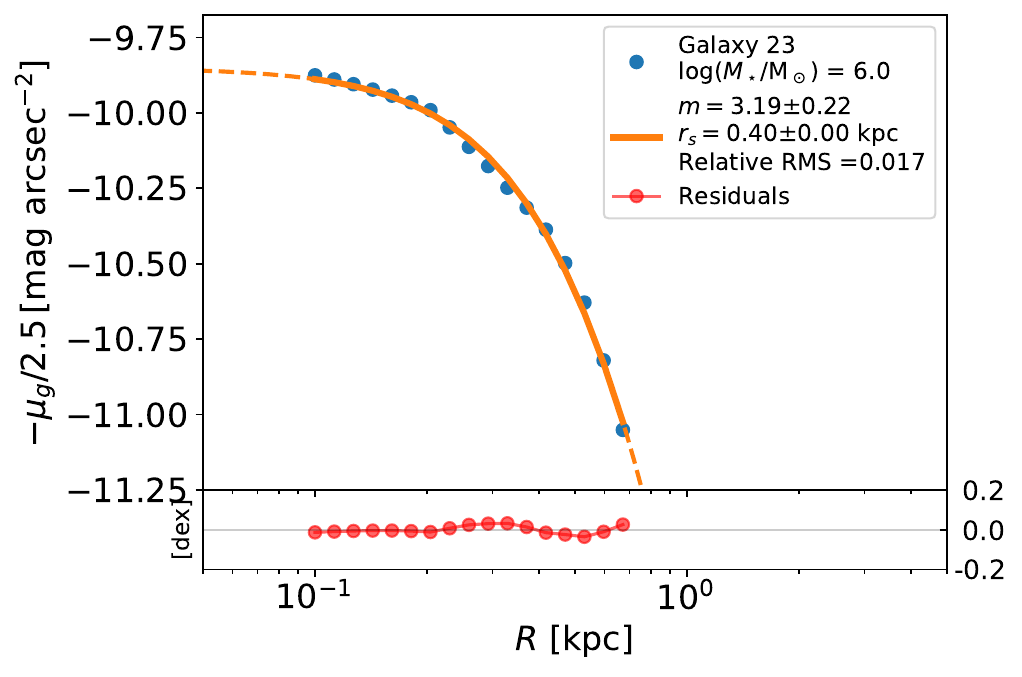}
\includegraphics[width=0.4\linewidth]{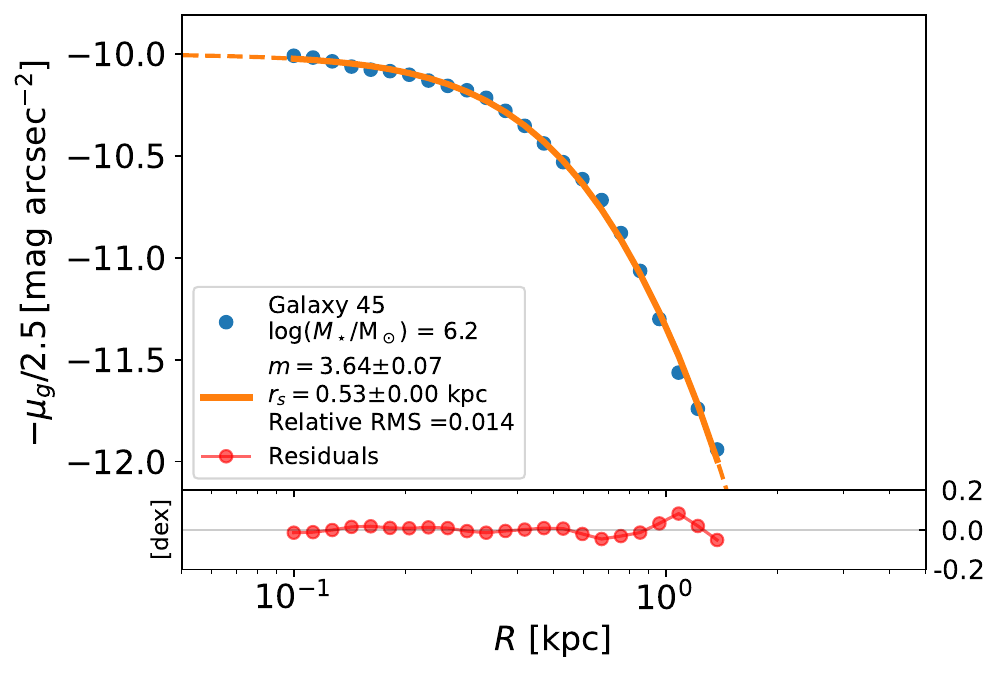}
\includegraphics[width=0.4\linewidth]{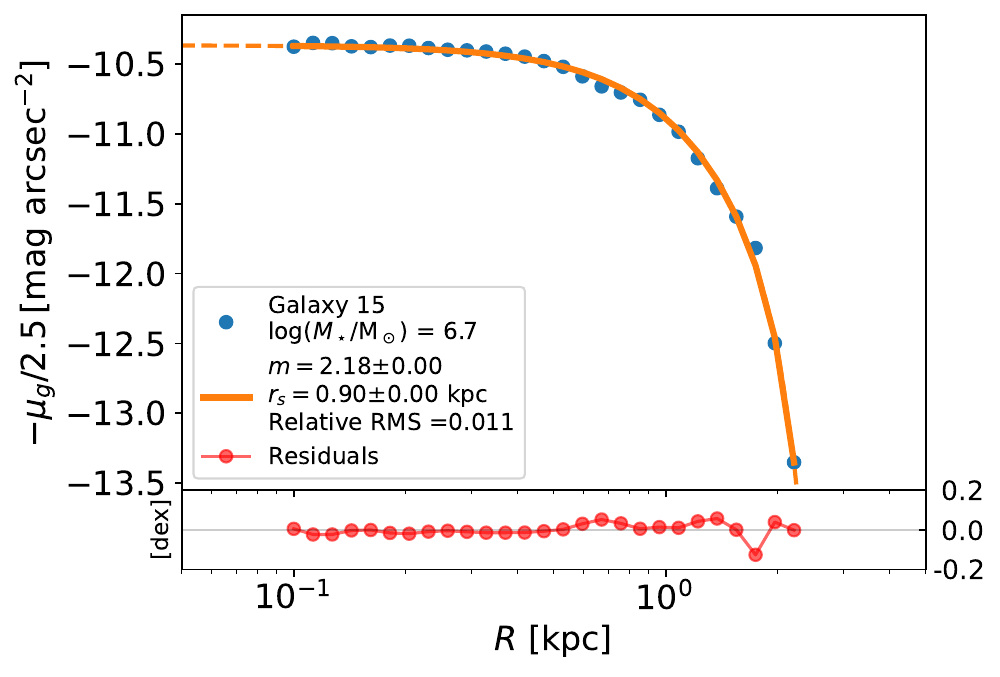}
\includegraphics[width=0.4\linewidth]{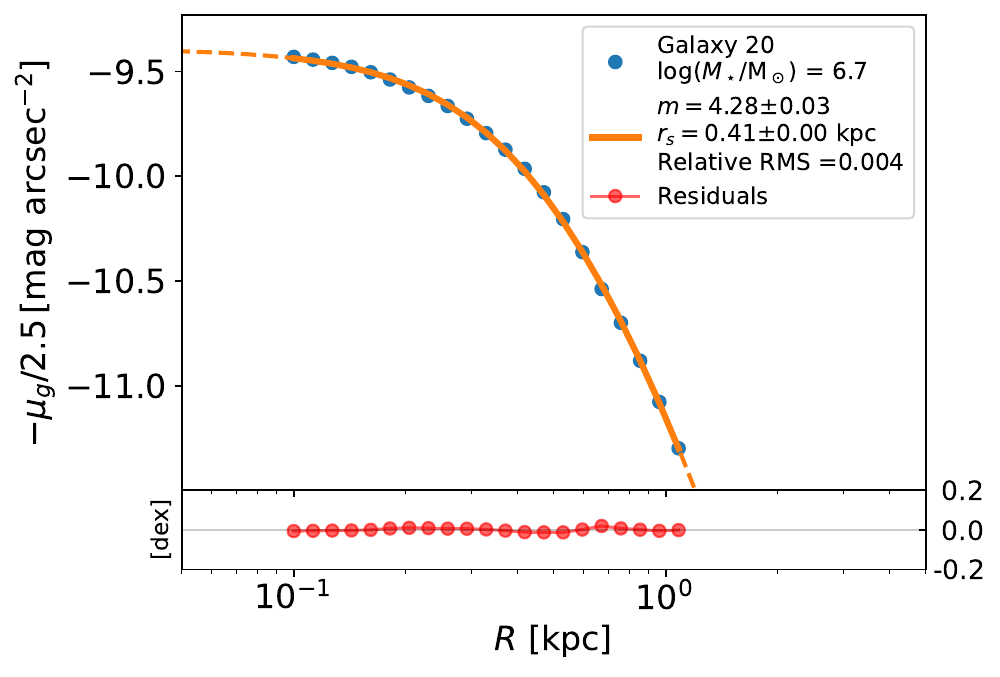}
\includegraphics[width=0.4\linewidth]{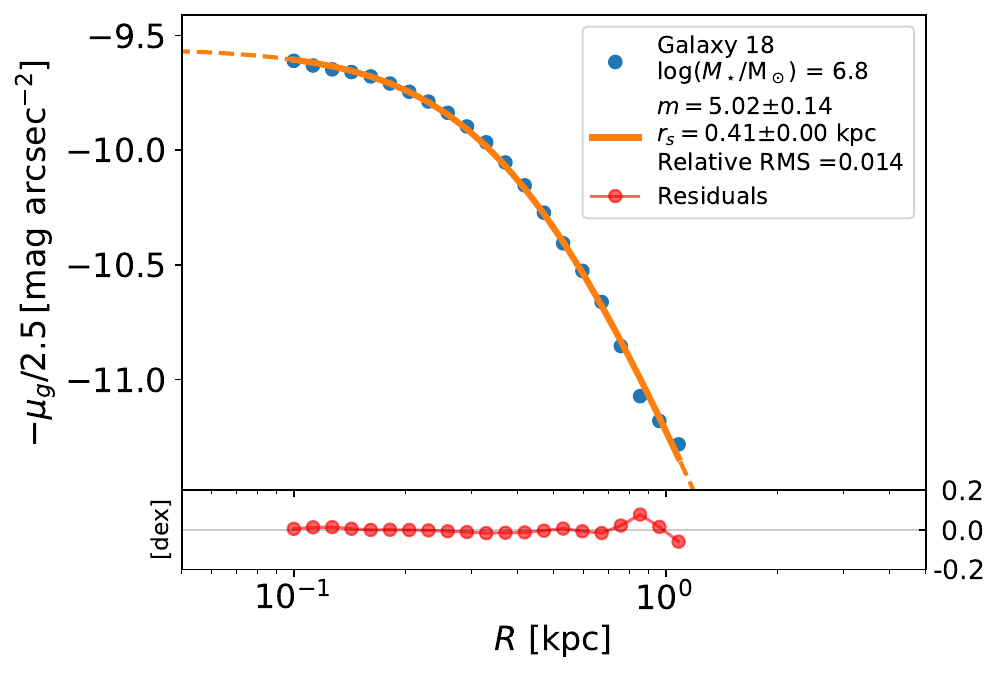}
\includegraphics[width=0.4\linewidth]{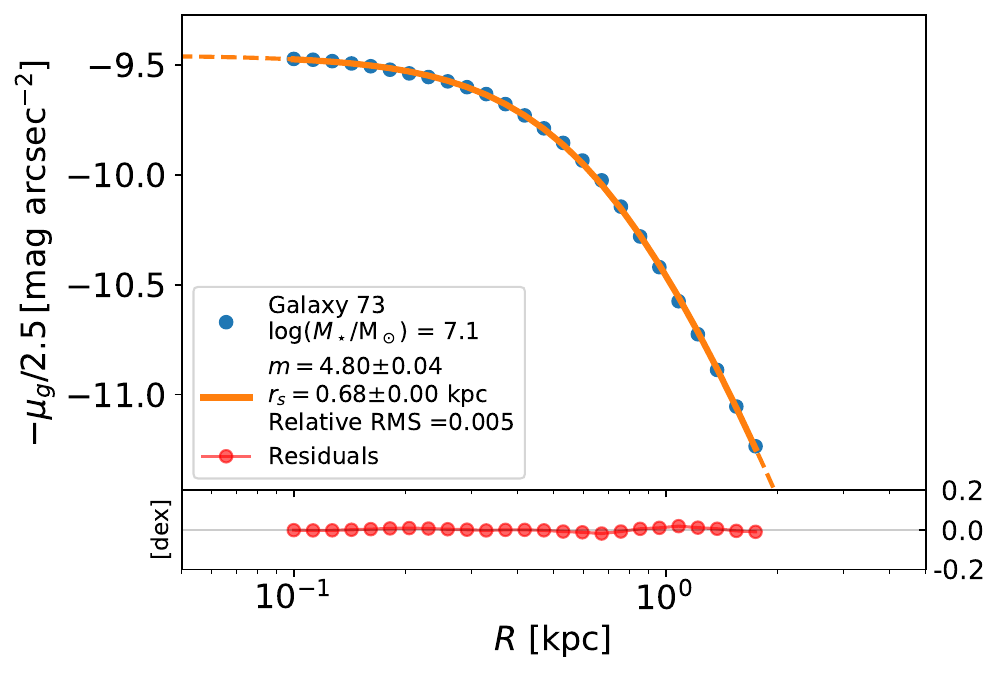}
\includegraphics[width=0.4\linewidth]{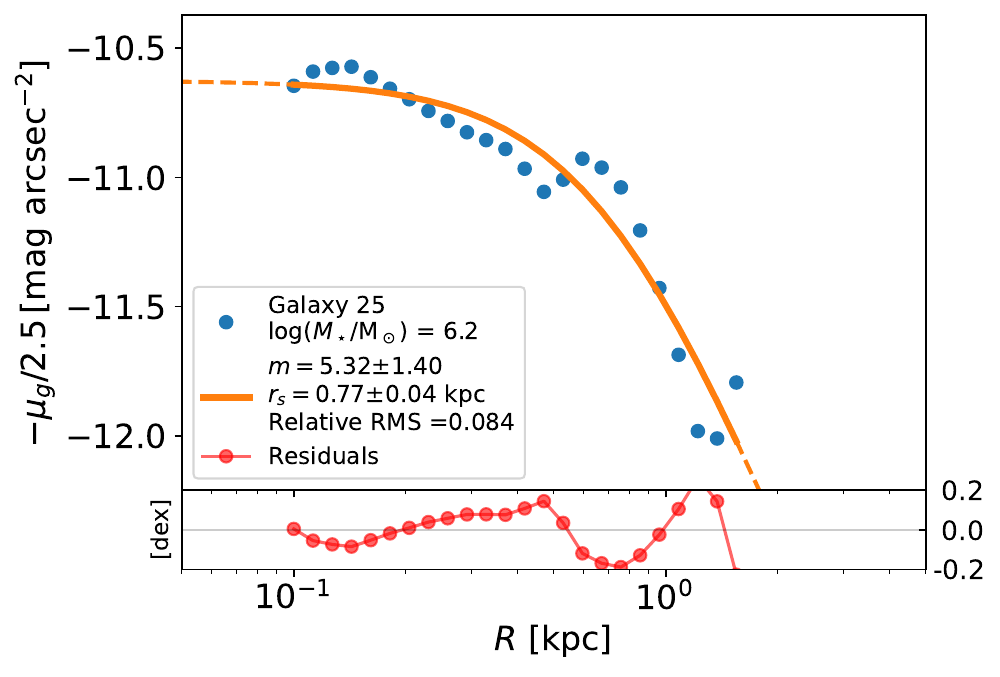}
\includegraphics[width=0.4\linewidth]{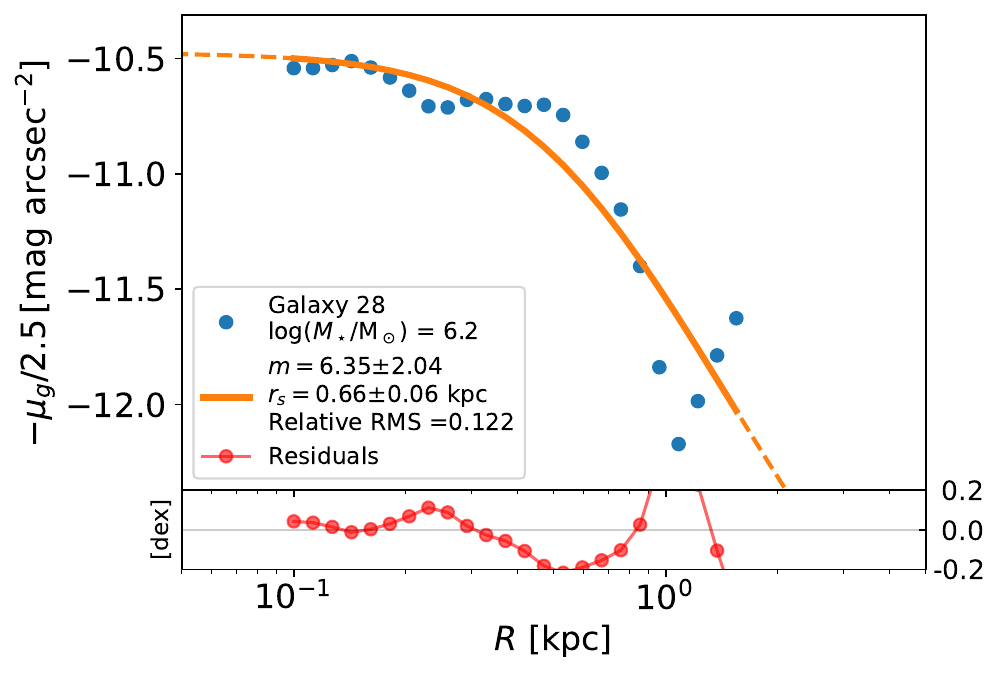}
\caption{Examples of projected polytrope fits to some of the profiles in Fig.~\ref{fig:raw_prof}. The first three rows contain god fits. The bottom row shows examples of bad fits, usually profiles with bumps resulting from deviations from axial symmetry of the original galaxies. The blue dots represent the observed points whereas the orange line shows the fit, which is solid or dashed depending on whether it is within or outside the range of fitted radii. The secondary sub-panel in each panel shows the residuals of the fit.
The insets provide information on the galaxy and the fit.}
  \label{fig:fitgood}
\end{figure*}
\section{Observed profiles and fits}\label{sec:observations}

This section describes the practical realization of steps number 1 and 2 in the procedure described in Sect.~\ref{sec:procedure}. 

The observed surface density mass distributions analyzed in the paper come from  \citet{2021ApJ...922..267C, 2022ApJ...933...47C}.  They provide the surface brightness  profile in the $g$ band ($\mu_g$) for a set of  dwarf satellites of Milky Way-like galaxies. They use photometry from various existing surveys including the  DESI Legacy Imaging Surveys \citep[][]{2019AJ....157..168D} and the Beijing–Arizona Sky Survey \citep{2017PASP..129f4101Z,2018ApJS..237...37Z}; see  \citet{2021ApJ...922..267C} for details. The galaxies are low mass dwarfs with stellar masses  between $10^6$ and $10^8\,{\rm M_\odot}$ as inferred from integrated photometry \citep{2013MNRAS.430.2715I}. The profiles are shown in Fig.~\ref{fig:raw_prof}, with the faintest signals in the outskirts reaching down to 30\,mag\,arcsec$^{-2}$. Even though the sky subtraction was carefully carried out  using ad-hoc custom-made procedures \citep[][]{2019ApJ...879...13C}, one cannot discard over or under corrections in  the faint end of the profiles ($\mu_g \gtrsim 29$\,mag\,arcsec$^{-2}$).

In principle, the EIM requires knowing the stellar mass density to be applied, but we use  $\mu_g$ directly because the shape of the DF is independent of a scaling factor in $\Sigma_o$ (Sect.~\ref{sec:procedure}) and
\begin{equation}
  \log\Sigma_o = -\mu_g/2.5 + \log\aleph = \log (\aleph F),
  \label{eq:defs1}
\end{equation}
with $\aleph$ the mass-to-light ratio. Obviously, for $\Sigma_o$ to scale as the surface brightness $F$, $\aleph$ has to be constant along the profile, which is a fair assumption for early-type galaxies, and the reason why we only use the early-type galaxies in  \citet{2021ApJ...922..267C} for our analysis\footnote{The early-type galaxies in the sample have colors consistent with stellar populations of the order of $6$\, Gyr \citep{2021ApJ...922..267C}. Assuming a 10\,\% age gradient between centers and outskirts (i.e., 0.6\,Gyr), one expects a gradient in $g-r$ around 0.03\,mag \citep[e.g.,][]{2003MNRAS.344.1000B} which translates into a change in $\log\aleph$ of 0.05 \citep[e.g.,][]{2013MNRAS.430.2715I}. Such change is negligible small compared with the observed range of $-\mu_g/2.5$ ($\gtrsim 2$; see Fig.~\ref{fig:fitgood}).}.
\citet{2021ApJ...922..267C} classified early and late-type galaxies by eyeball inspection of their morphology  so that smooth featureless objects are early types while objects with star-forming regions, blue clumps, or dust lanes are late types.
The original set contains 111 early-type galaxies, but it was  additionally purged to leave 93 as described in Sect.~\ref{sec:diagrams}.  

The volume density profiles required to compute the phase-space distribution function through the EIM (Eq.~[\ref{eq:ff1}]) are found by fitting various surface density profiles to the observed $ -\mu_g/2.5$. The fits are carried out using a least-squares minimization routine based on the Levenberg–Marquardt algorithm, reproducing the procedure employed in \citet{2021ApJ...921..125S}. We start off from a family of volume density profiles $\rho_\xi(r)$, with $\xi$ representing the free parameters that characterize the profile. Even when  $\rho_\xi(r)$ is analytical, its 2D projection is not, therefore,  Eq.~(\ref{eq:projection}) has to be evaluated numerically to carry out the fits. To speed up the procedure,  the surface densities required for fitting result from interpolating on a precomputed grid $\Sigma_\xi(R)$, related to $\rho_\xi(r)$ through Eq.~(\ref{eq:projection}). We employ a linear interpolation in as many dimensions as needed to account for the number of free parameters defining $\rho_\xi(r)$. The fits do not include the effect of the point-spread-function (PSF) on the shape of the synthetic profiles since the observed profiles are well resolved, even at the innermost radius  (100\,pc, Fig.~\ref{fig:raw_prof}). At the average distance of the observed galaxies ($\sim 7$\,Mpc), 100\,pc corresponds to 3\arcsec, which is significantly larger than the typical width of the PSF \citep[$\lesssim 1\arcsec$;][]{2021ApJ...922..267C}.

Figure~\ref{fig:fitgood} shows examples of this kind of fit to profiles in Fig.~\ref{fig:raw_prof}, where the density is fitted using polytropes. In this case the profiles depend on three parameters: a global density scaling, a spatial scaling $r_s$, and the polytropic index $m$ \citep[see,][]{2021ApJ...921..125S}.  Polytropes are relevant in this context because they represent self-gravitating systems of maximum Tsallis entropy  and so they describe the thermodynamical equilibrium of systems experiencing long-range forces \citep{1993PhLA..174..384P,2022Univ....8..214S}. Polytropes are know to describe the distribution of DM observed in dwarf galaxies  \citep{2020A&A...642L..14S}, the starlight observed in galaxies of various masses \citep{2021ApJ...921..125S},
as well as the DM distribution in numerical simulations of self-gravitating systems reaching thermodynamical equilibrium \citep{2021MNRAS.504.2832S}.   

\begin{figure}
\centering
\includegraphics[width=0.9\linewidth]{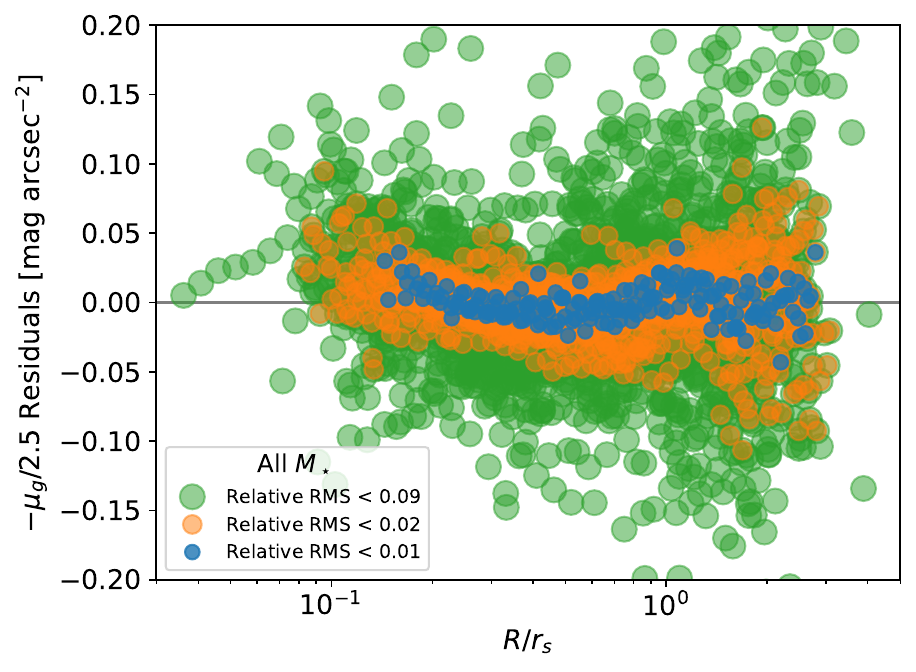}
\caption{Residuals of the fits to projected polytropes of the profiles in Fig.~\ref{fig:raw_prof}. The color code refers to the maximum Relative RMS defining each set of profiles. We define as good fits those represented in blue and orange.}
  \label{fig:eddington5_plot_a}
\end{figure}

Figure~\ref{fig:fitgood} shows both good (top three rows) and bad (bottom row) fits to some of the observed profiles.  The quality control parameter, called  ``Relative RMS'' in Fig.~\ref{fig:fitgood} and throughout the text, is defined as,
\begin{equation}
  {\rm Relative~RMS} = \frac{\rm RMS}{\max(\log F)-\min(\log F)},
  \label{eq:relative_rms}
\end{equation}
with RMS the root mean square of the residuals of the fit and $\log F$ the logarithm of the observed flux (Eq.~[\ref{eq:defs1}]). It represents the error in the fit relative to the amplitude of the observed signal. It is a good proxy for the quality of the fits so that 
$ {\rm Relative~RMS}  \simeq  0.02$  separates the good fits ($< 0.02$) form the fair fits ($> 0.02$) as judged rather arbitrarily from the eye-ball inspection of the individual fits (see Fig.~\ref{fig:eddington5_plot_a}). 
%
%

All polytropes have cores in the sense given in Eq.~(\ref{eq:strict}), therefore, to broaden the range of possibilities, we also try other more flexible density profiles, namely, the $\rho_{abc}$ profiles defined as
\begin{equation}
  \rho_{abc} = \frac{\rho_s}{x^c(1+x^a)^{(b-c)/a}},
  \label{eq:rhoabc}
\end{equation}
where $x=r/r_s$, and $\rho_s$ and $r_s$ are scaling constant in density and radius, respectively. These profiles are commonly used to model the density of baryons or DM \citep[e.g.,][]{1990ApJ...356..359H,2006AJ....132.2685M,2014MNRAS.441.2986D} and have the advantage of encompassing the iconic NFW profile  ($a=1, b=3, {\rm ~and~} c=1$) and the $m=5$ polytrope (a.k.a. Schuster-Plummer profile, with $a=2, b=5, {\rm ~and~} c=0$).
Moreover,  for $a=2$, $b=m$, and $c=0$, i.e., 
\begin{equation}
  \rho_{2m0} = \frac{\rho_s}{\left[1+(r/r_s)^2\right]^{m/2}},
  \label{eq:polapprox}
\end{equation}
it approximately accounts for the inner region of a polytrope of arbitrary index $m$ \citep[e.g.,][]{2022Univ....8..214S}.  The  $\rho_{abc}$ profile is more flexible than the polytropes because the inner and outer slopes can be fitted independently: the parameter $b$ encodes the outer logarithmic slope of the profile
\begin{equation}
  \lim_{r\to\infty}\frac{d\log\rho_{abc}}{d\log r} =-b, 
\end{equation}
whereas $c$ encodes the inner logarithmic slope (as in Eq.~[\ref{eq:innerlogs}]).
Examples of these alternative fits will be given later on in connection with the results (Sect.~\ref{sec:results}).

%
%
\begin{figure}
\centering
\includegraphics[width=0.95\linewidth]{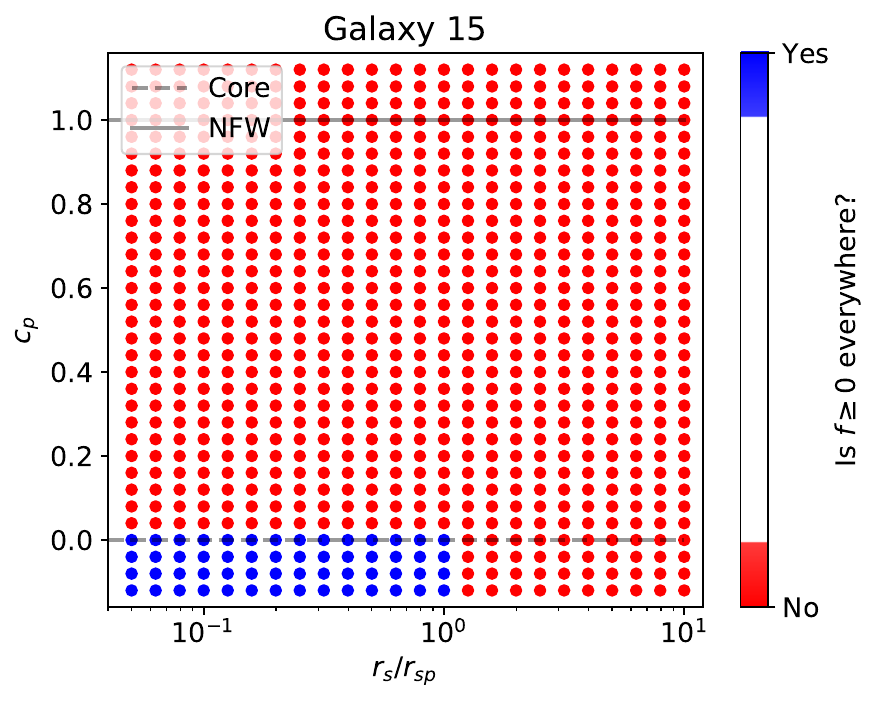}
\caption{Constraints on the underlying potential inferred from  the polytropic fit of a representative galaxy. The corresponding fit is shown in Fig.~\ref{fig:fitgood} (first column second row). Each potential is characterized by two parameters as shown in Eq.~(\ref{eq:rhoc}): the inner slope, $c_p$, and the global spatial scale, $r_{sp}$. Red dots correspond to $f < 0$ and so mark unphysical potentials whereas blue dots correspond to $f \geq 0$ and so trace potentials consistent  with the density profile. The parameter $r_s$ gives the spatial scale of the stellar light distribution. The horizontal lines define NFW potentials ($c_p=1$; the solid gray line) and cored potentials ($c_p=0$; the dashed gray line).}
  \label{fig:eddington1_plot_d0}
\end{figure}
\begin{figure}
\centering
\includegraphics[width=0.95\linewidth]{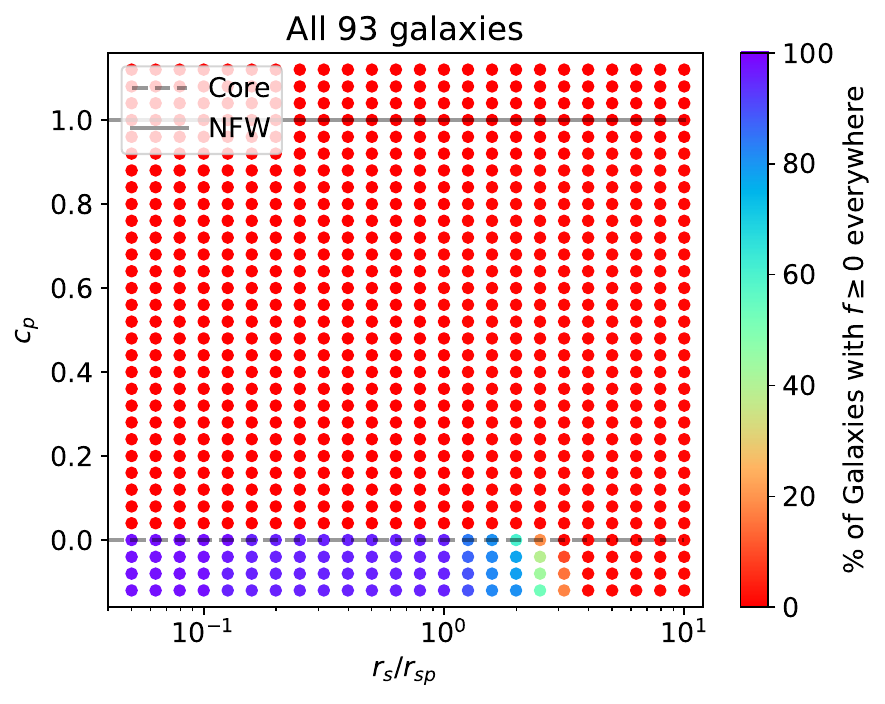}
\caption{Constraints on the underlying potential imposed by the set of polytropic fits like those in Fig.~\ref{fig:fitgood}. Every potential is characterized by $c_p$ and $r_{sp}$ and the color code shows the frequency of having $f \geq 0$ considering  all good and fair fits (93 galaxies). The whole region of $c_p> 0$ is discarded whereas $c_p\leq  0$ is permitted provided $r_{sp} \gtrsim  r_s/2$. The diagram does not change much when only good fits are considered   (Relative RMS $< 0.02$) or only low mass galaxies are selected  ($M_\star \leq 10^{7}\,{\rm M_\odot}$).
The parameter $r_s$ gives the spatial scale of the stellar light distribution. Horizontal lines defining NFW potentials ($c_p=1$; the solid gray line) and cored potentials ($c_p=0$; the dashed gray line) are included.}
  \label{fig:eddington1_plot_d}
\end{figure}
\section{Diagnostic diagrams}\label{sec:diagrams}

This section describes the practical realization of steps 3 and 4 in the procedure sketched out in Sect.~\ref{sec:procedure}. 

To constrain the properties of the gravitational potential, we compute for each observed profile the DFs in the phase-space corresponding to a battery of potentials. To make the interpretation intuitive, the potentials are described in terms the mass density profile that creates the potential  $\rho_p(r)$ (Eq.~[\ref{eq:pot_general}]).   In general the relation between $\rho_p$ and $\Psi$ is not analytical and has to be computed numerically. For $\rho_p$ we use  a particular type of $\rho_{abc}$ function (Eq.~[\ref{eq:rhoabc}]), 
\begin{equation}
  \rho_{p}(r) = \frac{\rho_{sp}}{x^{c_p}(1+x^{2-{c_p}})^{(5-3{c_p})/(2-{c_p})}},
\label{eq:rhoc}
\end{equation}
that scans  seamlessly from a Schuster-Plummer profile to a NFW profile when $c_p$ goes from 0 to 1. Thus, the potentials in the battery are defined in terms of the two free parameters that define $\rho_p$, i.e.,  $c_p$ and $r_{sp}$. (The scaling factor $\rho_{sp}$ is irrelevant for the EIM, as pointed out in Sect.~\ref{sec:eddington_summary}.) The key question of the analysis is knowing whether the DF resulting from the observed profile and each potential is or not negative. The answer to the question can be summarized in the kind of diagnostic diagram shown in Fig.~\ref{fig:eddington1_plot_d0}. Red dots correspond to $f < 0$ and so mark unphysical potentials whereas blue dots corresponds to $f \geq 0$ and so trace gravitational potentials consistent  with the density profile.

Figure~\ref{fig:eddington1_plot_d0} shows a typical example corresponding to galaxy \#\,15, with good polytropic fit in Fig.~\ref{fig:fitgood}. Here an throughout the paper $f$ is computed using the methods and tools developed by  \citet{2023ApJ...954..153S}. These tools only work for $\rho_{abc}$ profiles, therefore, for this calculation polytropes were approximated using Eq.~(\ref{eq:polapprox}). Note that all  $c_p > 0$ are discarded in Fig.~\ref{fig:eddington1_plot_d0}. All potentials have to derive from a cored density profile to be consistent with the fitted polytropes. This comes with no surprise since all polytropes have a core \citep[e.g.,][]{2020A&A...642L..14S} and we proved that a core in stars requires a core in the potential for spherically symmetric systems and isotropic velocity distributions (see details in Sect.~\ref{sec:eddington_summary}). In addition, some of the models with $r_s \geq r_{sp}$ are also incompatible. We also knew from \citet{2023ApJ...954..153S} of the inconsistency of potentials with characteristic radii ($r_{sp}$) significantly smaller than the stellar core radii ($r_s$). Perhaps more surprising is the fact that no incompatibility is found for some potentials with $c_p < 0$, i.e., where the total density goes to cero when $r\to 0$. These potentials are included in the battery for comprehensiveness although they are expected to be Rayleigh–Taylor unstable \citep[e.g.,][]{2022ApJ...940...46S} and so of difficult practical realization.  Once again, \citet{2023ApJ...954..153S} showed that in this case ($c=0$, $a=2$, and $c_p< 0$) the derivative $d\rho/d\Psi > 0$ when $r\to 0$ and so there is no need for $f$ to be negative (see their Appendix~E).

From the 111 galaxies in Fig.~\ref{fig:raw_prof},  we discard those having bad fits (like those of the bottom row in Fig.~\ref{fig:fitgood}), and then are left with 93 good and fair fits.  For every one of these profiles, we have a diagnostic diagram like the example shown in Fig.~\ref{fig:eddington1_plot_d0}. We construct a second kind of diagnostic diagram to show the constraints on the underlaying potential imposed for the whole set of observed profiles. For every potential, characterized by $c_p$ and $r_s/r_{sp}$, we count the frequency of $f \geq 0$ considering all the observed galaxies. The result is shown in Fig.~\ref{fig:eddington1_plot_d}. As we knew, the whole region of $c_p> 0$ is discarded by all (100\,\%) of the fits. The region with $c_p\leq  0$ is permitted provided $r_{sp} \gtrsim  r_s/2$. The results do not change much when only the good fits are used (Relative RMS $< 0.02$) or when only the low mass galaxies are considered ($M_\star < 10^7\,{\rm M_\odot}$).

Formal error bars can be assigned to this kind of diagnostic diagram as follows. It is based on counting the number of galaxies that, for the same potential, do or do not have a particular property (i.e., whether or not $f \geq 0$ everywhere). Thus, the actual number of counts should follow a binomial distribution. Using this idea, we work out in Appendix~\ref{app:errors} the error bars to be expected. They are largest when the percentage is 50\,\%, and tend to zero at 0\,\% and 100\,\%. The actual error depends on the number of galaxies employed in the diagram. For $\sim 50$ galaxies (typical number of good galaxies or low-mass galaxies in our sample), it is smaller than 6\,\% (Fig.~\ref{fig:appa}).
%

%
\section{Results}\label{sec:results}

Armed with the fits described in Sect.~\ref{sec:observations} and the diagnostic diagrams of Sect.~\ref{sec:diagrams}, here we describe the constraints on the potentials set by the observed surface brightness  profiles. The degrees of freedom in the fits increase from polytropes (Sect.~\ref{sec:pp_results}) to variable inner slope (Sect.~\ref{sec:intermediate}) to allow both the inner and outer slopes to vary simultaneously (Sect.~\ref{sec:full}). 
%
%
\begin{figure}
\centering
\includegraphics[width=0.9\linewidth]{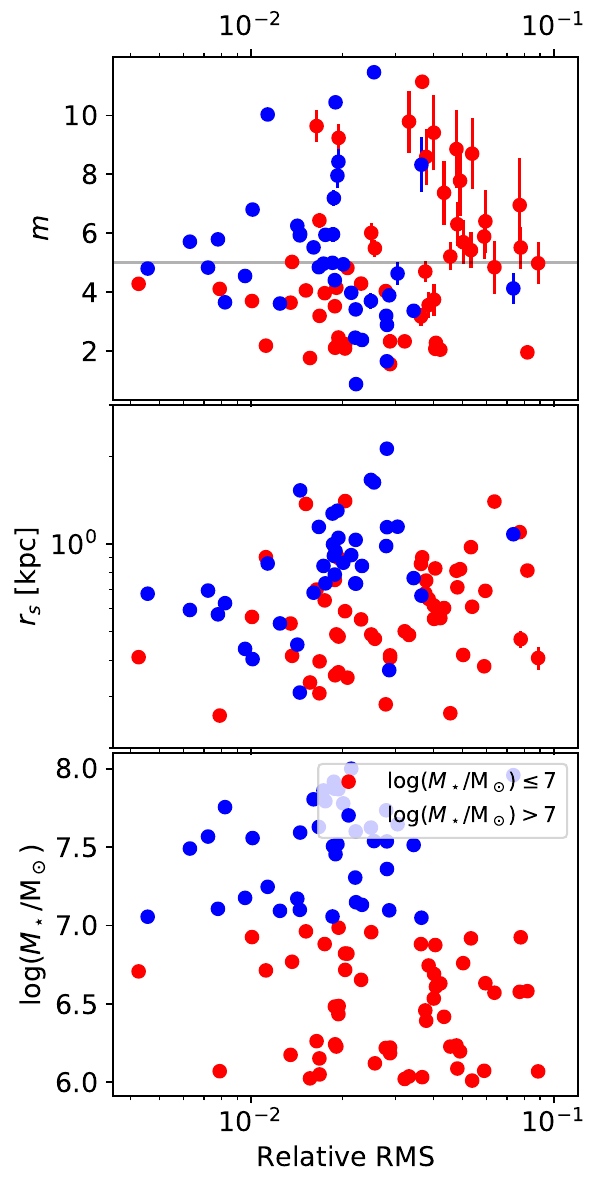}
\caption{Quantitative summary with the results of the polytropic fit to the profiles in Fig.~\ref{fig:raw_prof}. Low ($\leq  10^7\,{\rm M_\odot}$) and high  ($> 10^7\,{\rm M_\odot}$) stellar mass galaxies are represented with a different color as indicated in the bottom panel.  
  Top panel: dependence of the inferred polytropic index $m$ on the quality of the fits, as parameterized by the Relative RMS. There seems to be a trend for the best quality fits to have $m\sim 5$ (marked with the gray solid horizontal line).
  Middle panel: scatter plot with the size of the core versus the goodness of the fit.
  Bottom panel: scatter plot with the stellar mass versus the goodness of the fit. More massive galaxies lead to slightly better fits. 
}
  \label{fig:eddington1_plot_new_b}
\end{figure}
\begin{figure}
\centering
\includegraphics[width=0.9\linewidth]{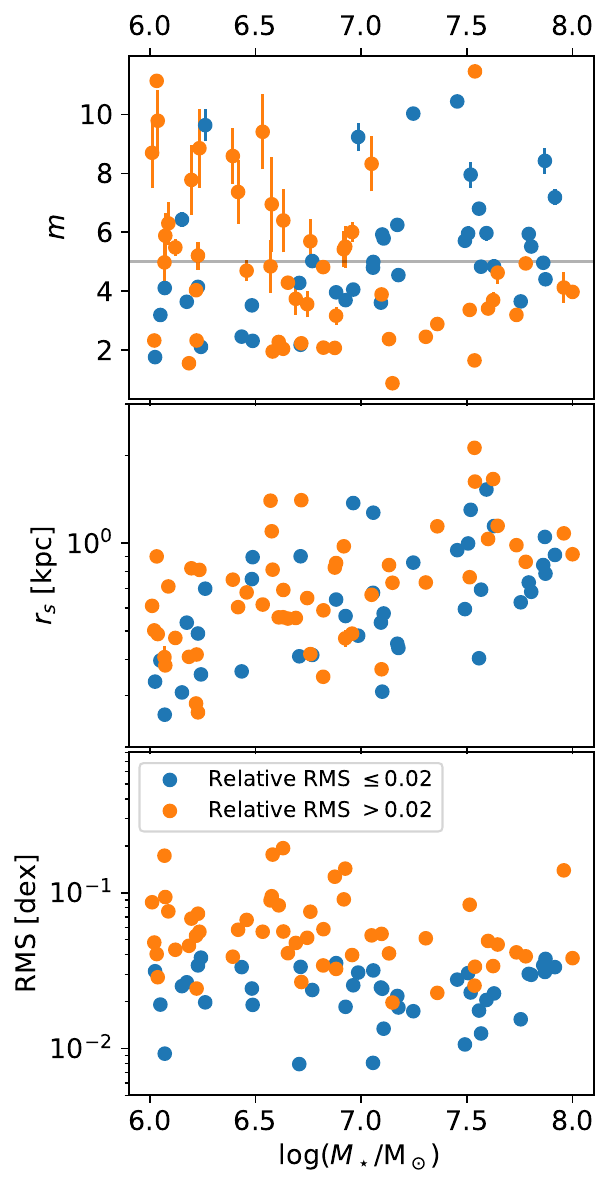}
\caption{Quantitative summary with the results of the polytropic fit to the profiles in Fig.~\ref{fig:raw_prof}. Medium (Relative RMS $> 0.02$) and high  ($\leq 0.02$) quality fits are represented with different colors as indicated in the bottom panel. 
  Top panel: scatter plot of the polytropic index $m$ as a function the stellar mass of the galaxy. 
  Middle panel:  variation of the core radius with stellar mass.  More massive galaxies tend to have larger cores.
  Bottom panel: scatter plot with the variation of the absolute RMS with stellar mass.  Good fits (blue dots) have absolute RMS $\lesssim  0.03$\,dex.
  }
  \label{fig:eddington1_plot_new_c}
\end{figure}
%
\begin{figure*}
\centering
\includegraphics[width=0.4\linewidth]{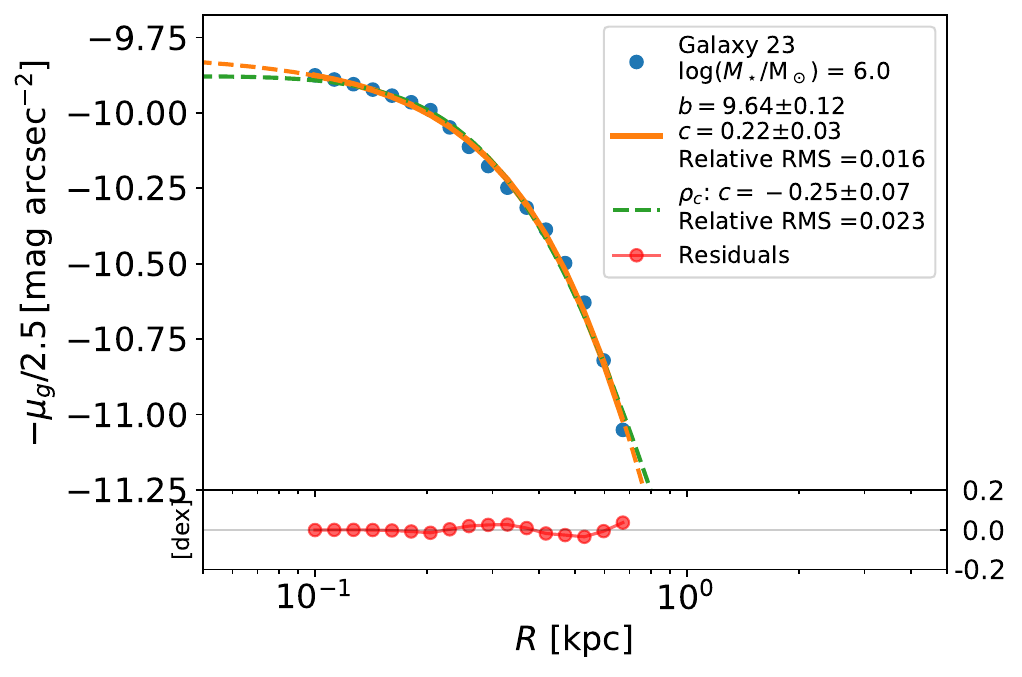}
\includegraphics[width=0.4\linewidth]{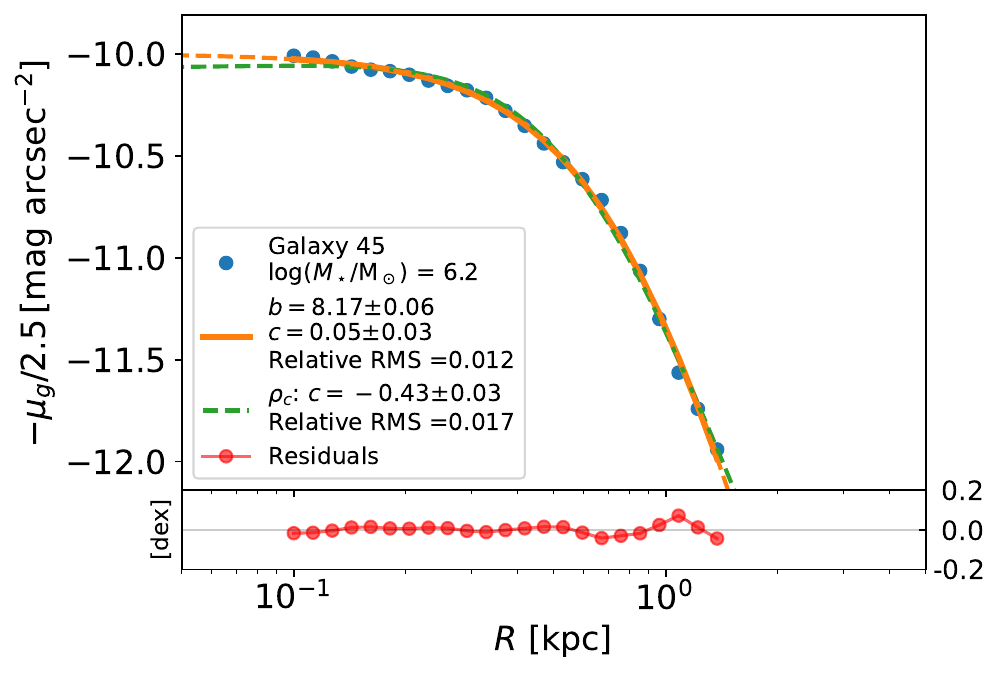}
\includegraphics[width=0.4\linewidth]{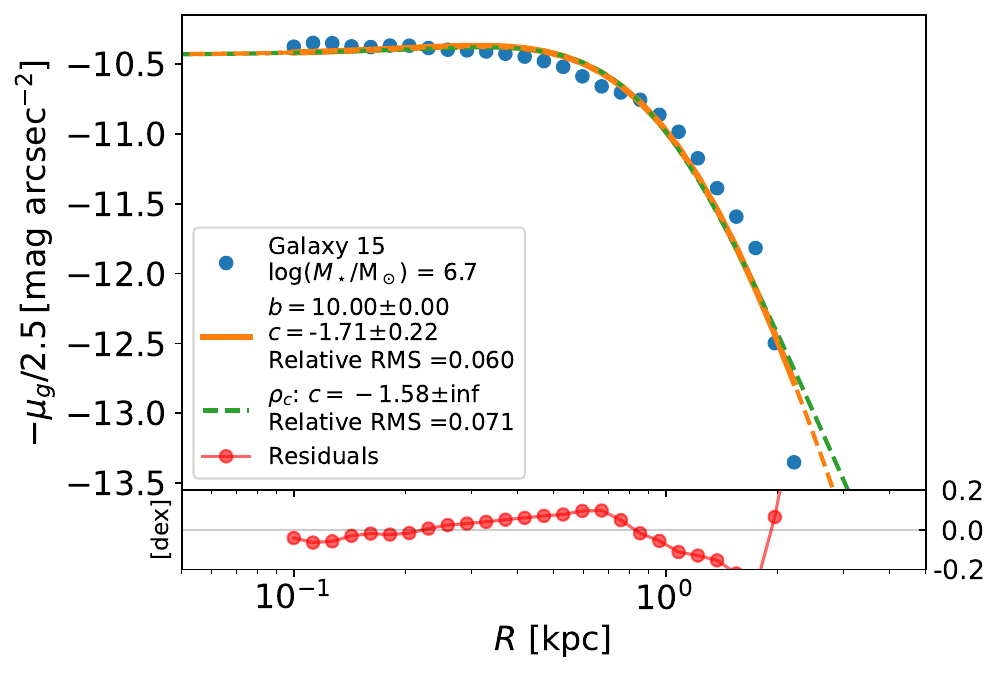}
\includegraphics[width=0.4\linewidth]{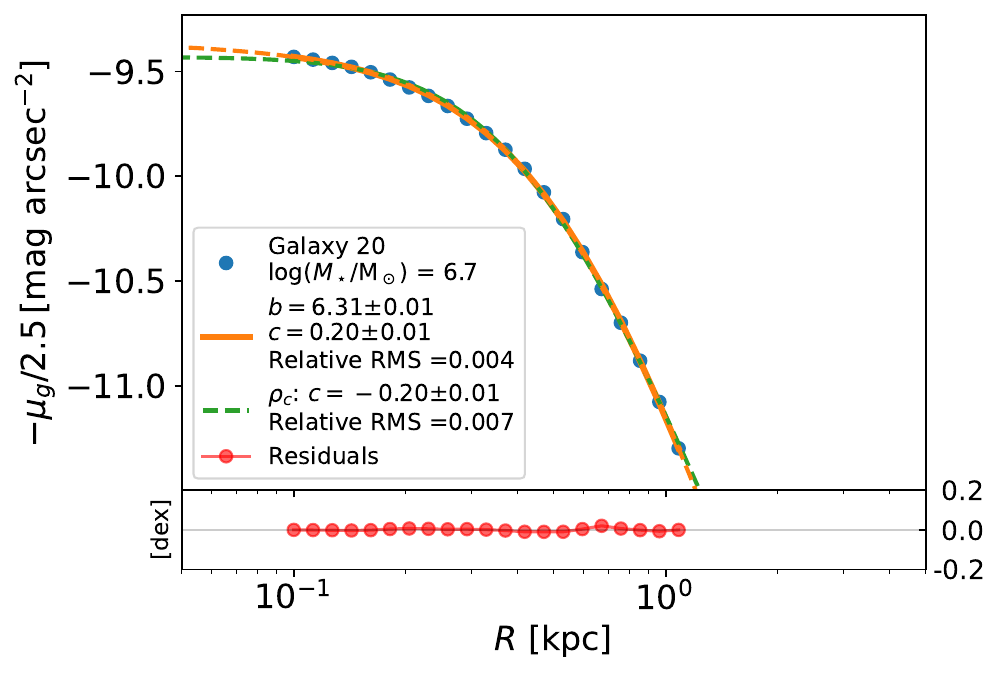}
\includegraphics[width=0.4\linewidth]{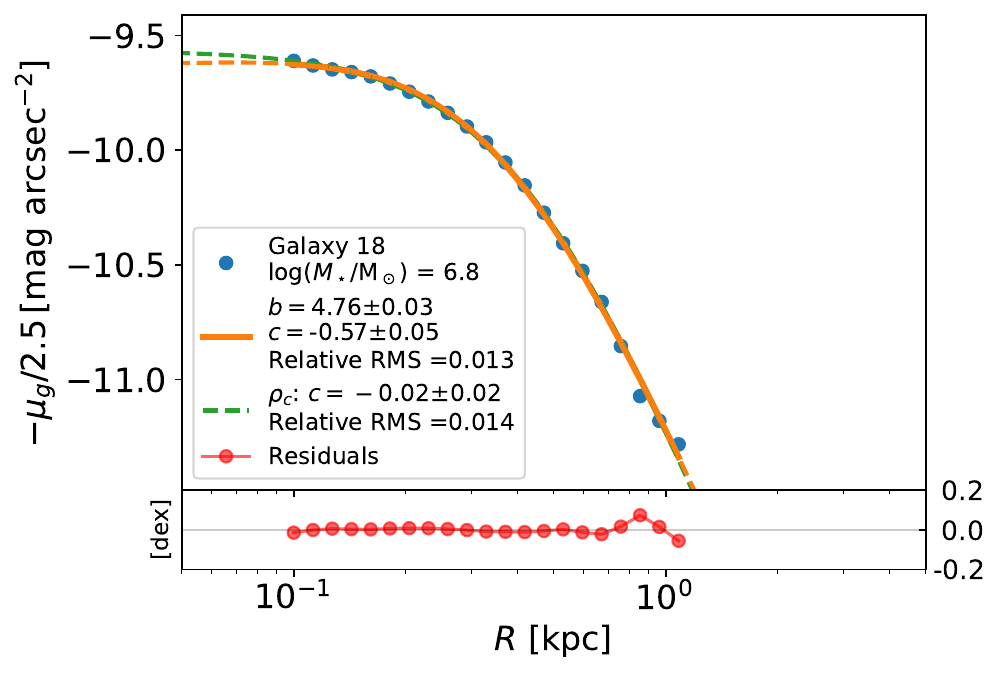}
\includegraphics[width=0.4\linewidth]{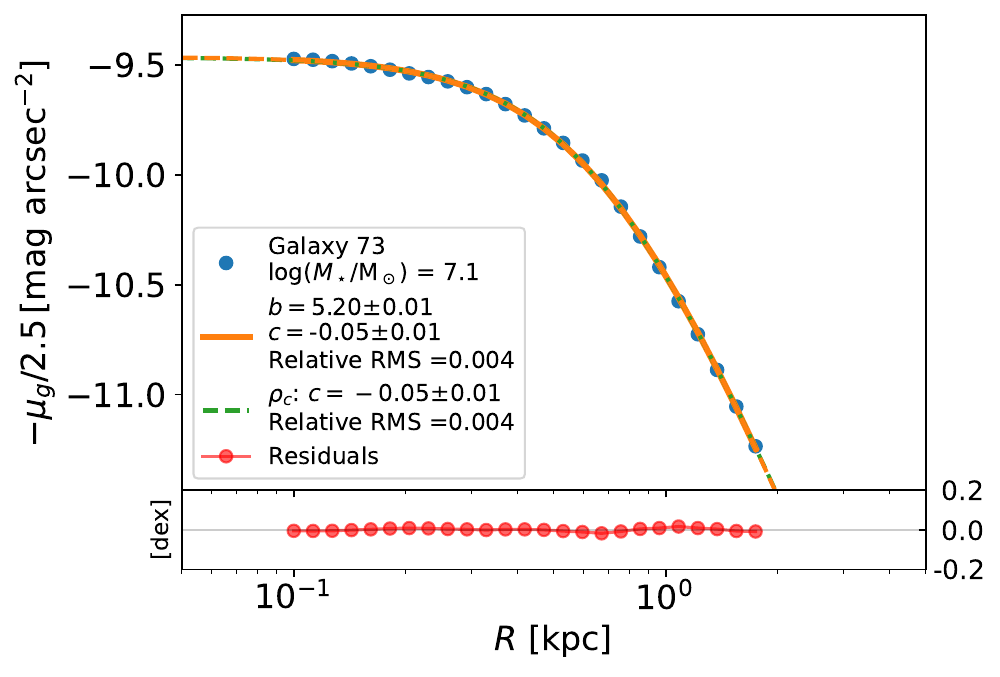}
\includegraphics[width=0.4\linewidth]{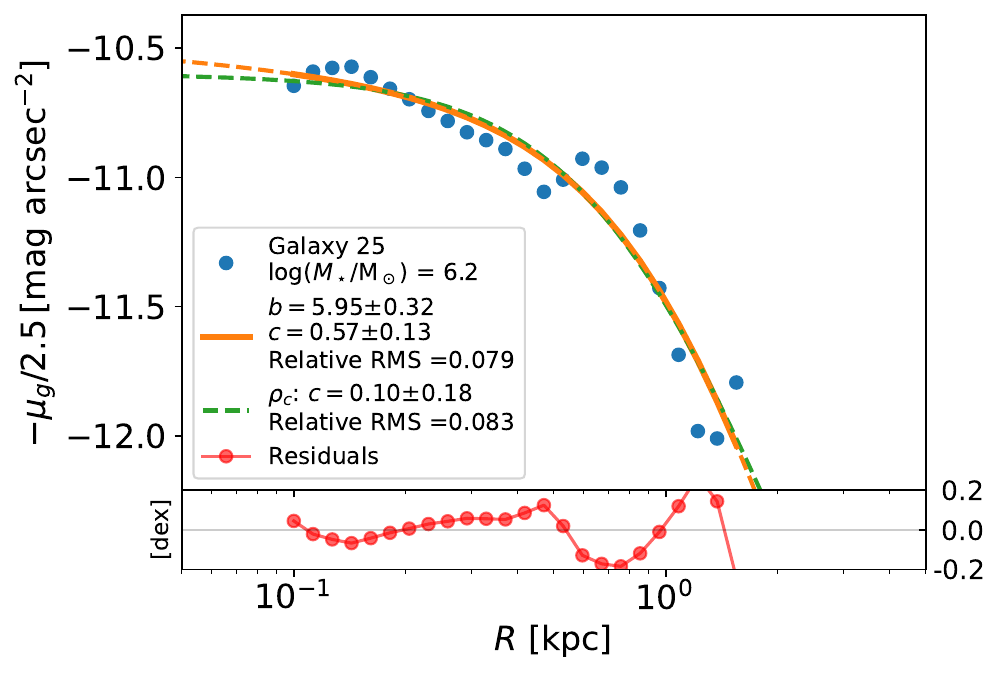}
\includegraphics[width=0.4\linewidth]{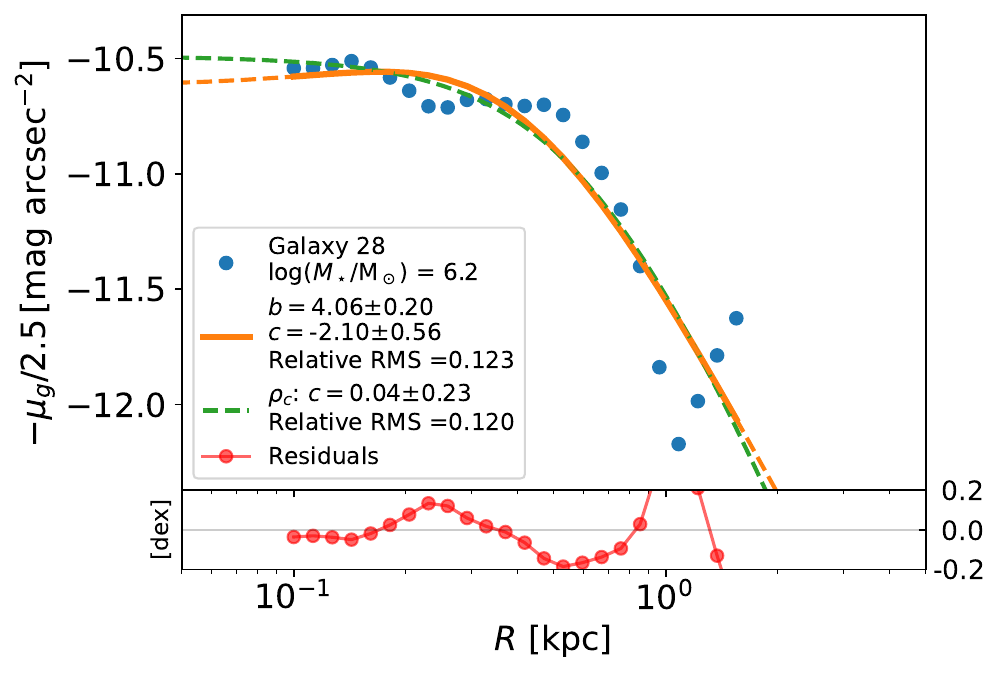}
\caption{Fits similar to Fig.~\ref{fig:fitgood} but allowing the inner slope of the stellar mass profile to vary  (green dashed lines), and the inner and outer slopes to vary independently (orange lines). The density used for fitting was assumed to be $\rho_{abc}$ in Eq.~(\ref{eq:rhoabc}) with different bonds between $a$, $b$, and $c$ for the green and the orange lines. 
  See the main text and the caption of Fig.~\ref{fig:fitgood} for further details.}
  \label{fig:fitgood_2bc}
\end{figure*}
\begin{figure}
\centering
\includegraphics[width=0.9\linewidth]{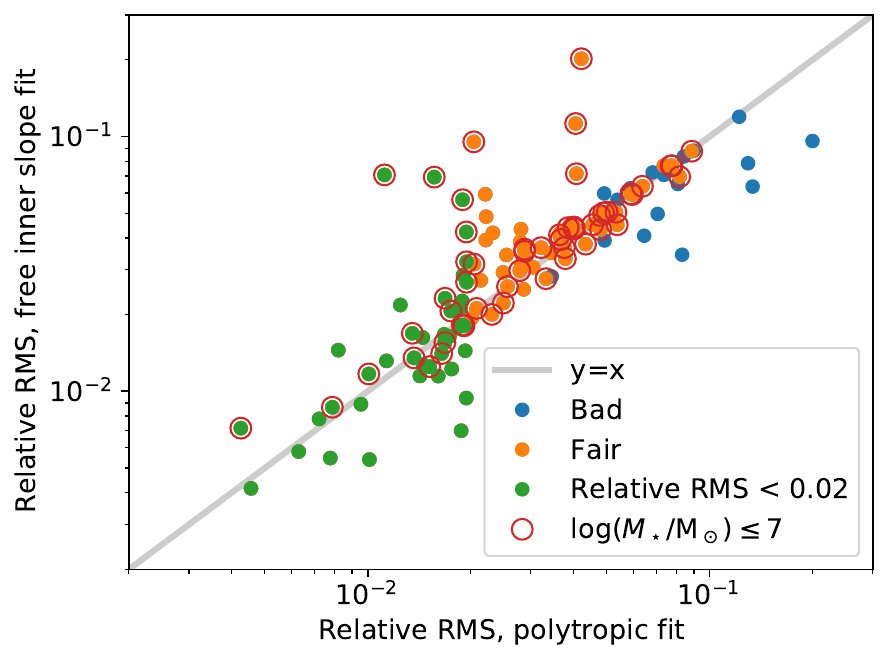}
\caption{Scatter plot of the Relative RMS resulting from variable inner slope fits and from polytropic fits. Different colors designate different quality of the polytropic fits (see the inset). The points corresponding to low mass galaxies are  encircled in red. The absolute number of good polytropic fits is larger as is also the relative fraction of good over bad fits.}
  \label{fig:eddington3_plot}
\end{figure}

\subsection{Polytropic fits, with zero inner slope}\label{sec:pp_results}
Examples of polytropic fits are shown in Fig.~\ref{fig:fitgood}. The quality of these fits is quantified using the Relative RMS (Eq.~[\ref{eq:relative_rms}]) and overall, the quality is satisfactory (see the residuals in Fig.~\ref{fig:eddington5_plot_a}). As we mentioned in Sect.~\ref{sec:diagrams}, from the 111 original profiles in Fig.~\ref{fig:raw_prof}, 18 are discarded because the profile is not smooth and the fits are bad (like those of the bottom row of Fig.~\ref{fig:fitgood}). From the remaining 93, 41 are good  (44\,\%, with Relative RMS$< 0.02$) and the rest fair (64\,\%).

The results of the fits are summarized in Figs.~\ref{fig:eddington1_plot_new_b} and \ref{fig:eddington1_plot_new_c}. The top panel of  Fig.~\ref{fig:eddington1_plot_new_b} shows the dependence of the inferred polytropic index $m$ as a function the quality of the fits parameterized by  the Relative RMS. There seems to be a trend for the best quality polytropic fits to have $m\sim 5$, a behavior also found in other galaxy samples \citep{2021ApJ...921..125S}. This trend does not depend on the stellar mass, divided in low ($\leq$\,$10^7 M_\odot$) and high ($>$\,$10^7 M_\odot$) mass bins for the plot.  The middle  panel of  Fig.~\ref{fig:eddington1_plot_new_b} contains the size of the core versus the goodness of the fit. No trend is found. The bottom  panel of  Fig.~\ref{fig:eddington1_plot_new_b} shows the stellar mass versus the goodness of the fit, and there is a hint of trend so that fits are slightly better for the more massive galaxies.
This effect may be an observational bias since more massive objects tend to be brighter and thus present better observations that leave smaller residuals.
The top panel of  Fig.~\ref{fig:eddington1_plot_new_c} shows the polytropic index as a function the stellar mass of the galaxy.  Overall, there is no clear trend of $m$ with  $\log M_\star$, which may be due to the fact that the lowest mass galaxies happen to have the worst fits where $m$ is more uncertain. However, the good fits seem to indicate a trend for $m$ to grow with increasing $M_\star$ (the blue symbols in Fig.~\ref{fig:eddington1_plot_new_c}, top panel).
The middle panel of Fig.~\ref{fig:eddington1_plot_new_c} displays the variation of the core radius with stellar mass.  More massive galaxies tend to have larger cores. This trend does not depend on the quality of the fits, represented with two colors depending on whether the Relative RMS is smaller or larger than 0.02.    The bottom panel of Fig.~\ref{fig:eddington1_plot_new_c} shows the variation of the absolute RMS with stellar mass.  Note that the good fits have an absolute RMS better than 0.03\,dex, equivalent to 7\,\%\ in  flux.

With the polytropic fits described above, we constructed the diagnostic diagrams shown in Figs.~\ref{fig:eddington1_plot_d0} and \ref{fig:eddington1_plot_d}. They were discussed as examples to present the diagnostic diagrams in Sect.~\ref{sec:diagrams}, and there is little else to add except to emphasize that all kinds of cuspy potentials are discarded ($c_p > 0$) and the characteristic radius of the potential cannot be significantly smaller than the radius of the stellar distribution  ($r_{sp} \gtrsim  r_s/2$). The first result was to be expected since given the hypotheses made in our application of the EIM, a core in stars necessarily needs of a core in the potential, and all polytropes are cored profiles independently of their index.   

%
\begin{figure}
\centering
\includegraphics[width=0.9\linewidth]{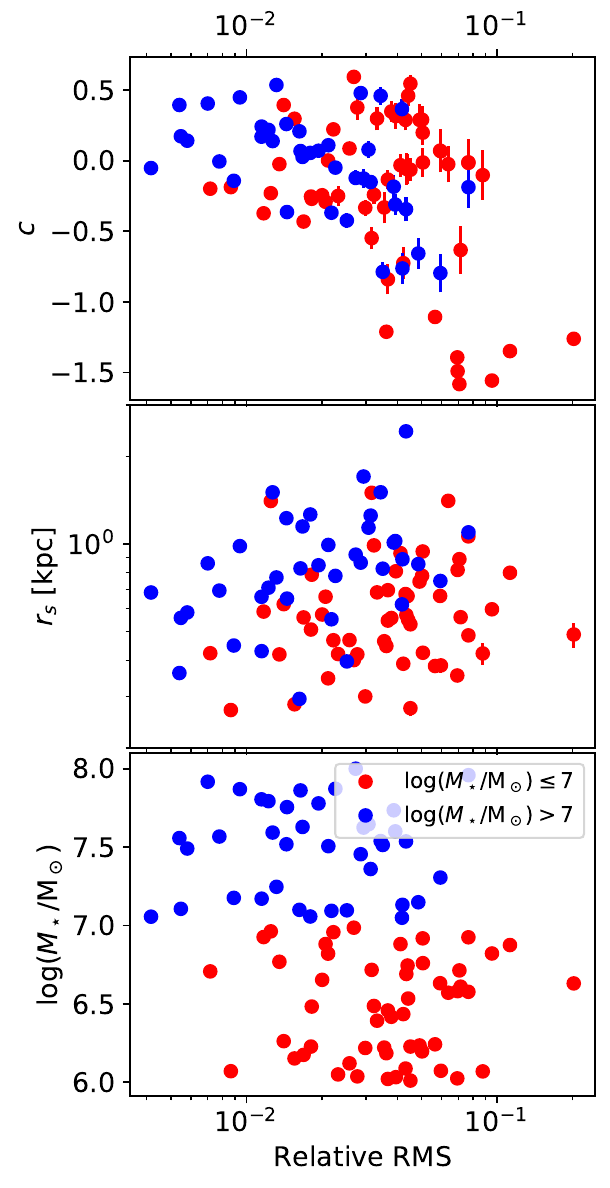}
 \caption{Quantitative summary with the results of the variable inner slope fit. Except for the inner slope $c$, the rest is similar to Fig.~\ref{fig:eddington1_plot_new_b} describing  polytropic fits. Note how $c$ is often negative, a trend enhanced as the quality of the fits worsens.}
  \label{fig:eddington2_plot_new_b}
\end{figure}
\begin{figure}
\centering
\includegraphics[width=0.9\linewidth]{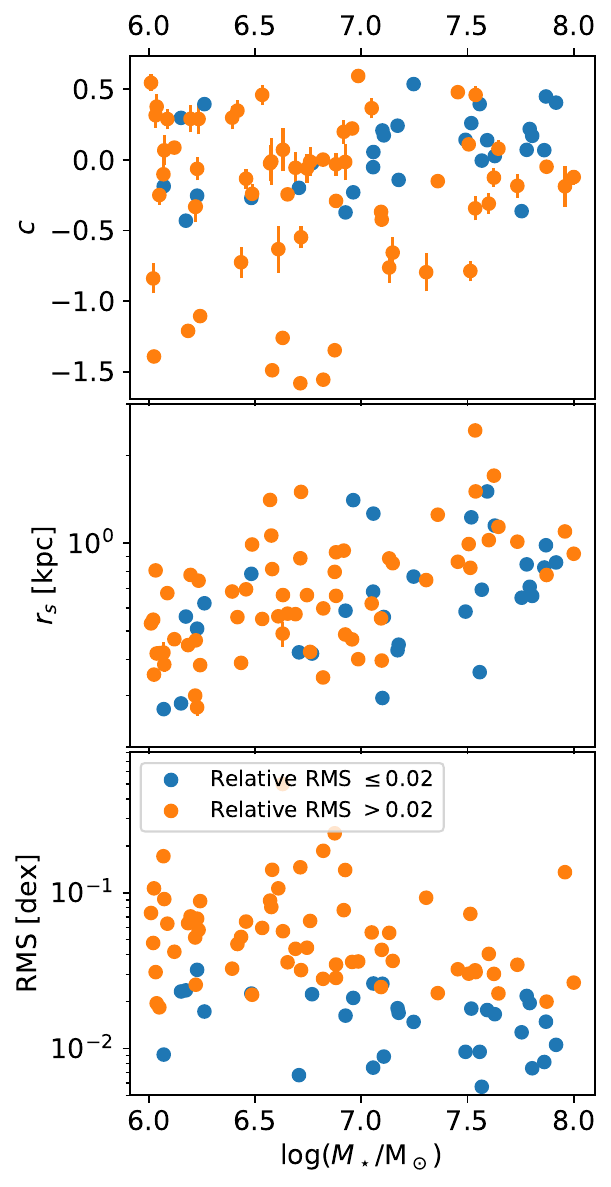}
  \caption{Quantitative summary with the results of the variable inner slope fit. Except for the inner slope $c$ replacing the polytropic index, the figure is similar to Fig.~\ref{fig:eddington1_plot_new_c} describing the polytropic fits.}
  \label{fig:eddington2_plot_new_c}
\end{figure}
\begin{figure}
\centering
\includegraphics[width=0.95\linewidth]{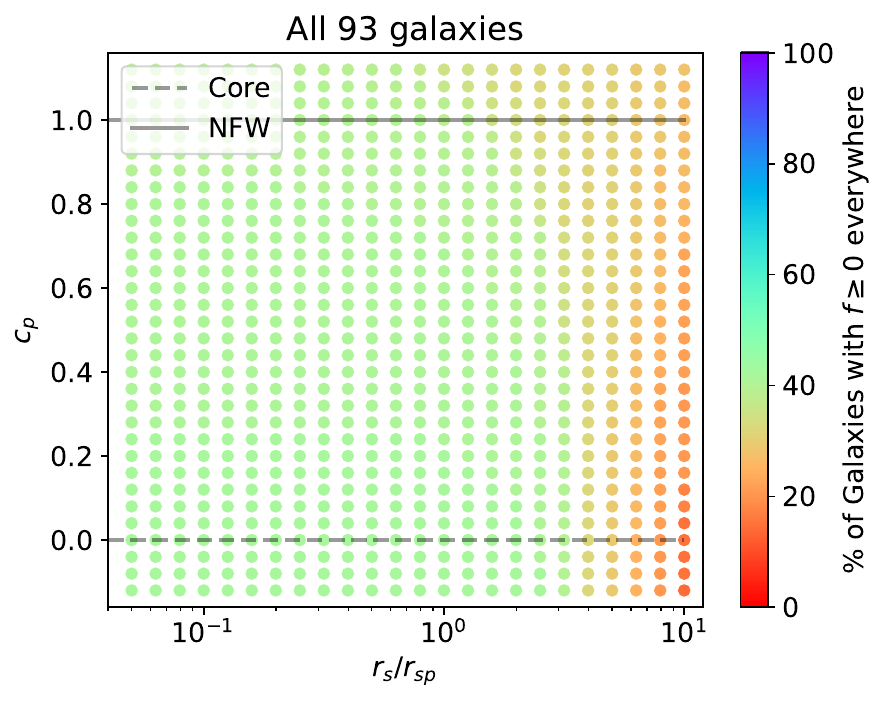}
\includegraphics[width=0.95\linewidth]{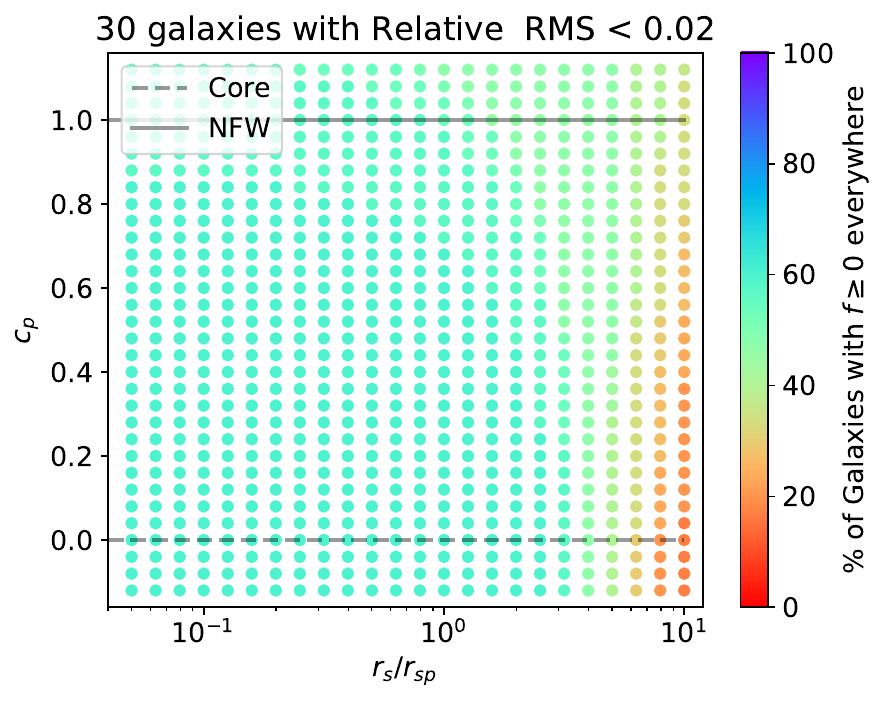}
\includegraphics[width=0.95\linewidth]{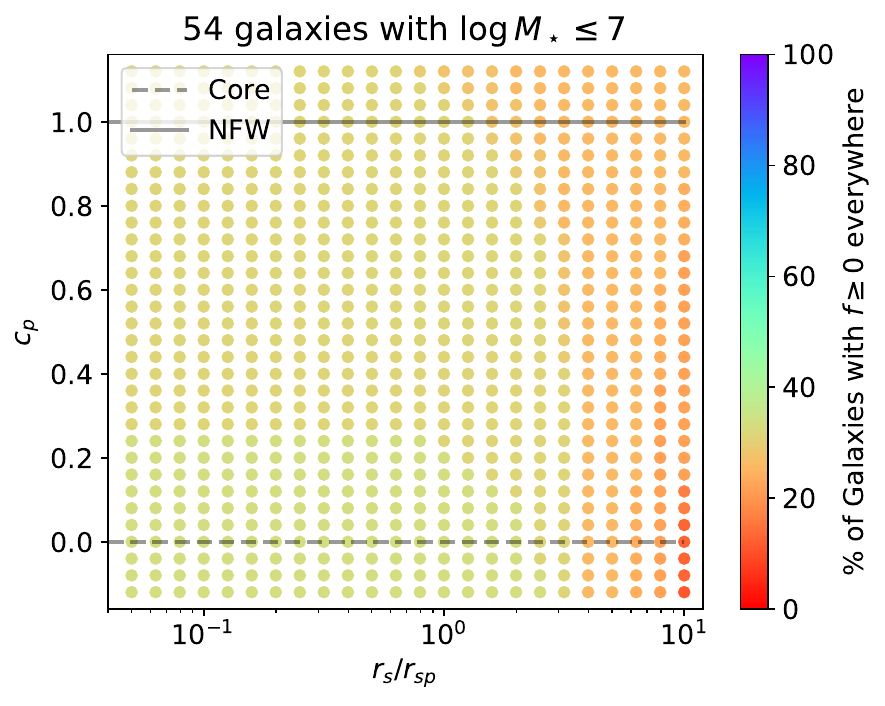}
\caption{Constraints on the underlying gravitational potential imposed by the variable inner slope fits. Top panel: diagnostic diagram, equivalent to  Fig.~\ref{fig:eddington1_plot_d} for the polytropic fits, which includes all galaxies together. Note that around half of the galaxies have unphysical $f <  0$ irrespectively of the inner slope of the potential ($c_p$), a percentage that increases as $r_s/r_{sp}$ increases. This is mostly due to the inner slope $c$ being negative for a number of galaxies. Middle panel: similar to the top panel but including only good fits (Relative RMS $<0.02$). Note how the fraction of unphysical potentials with $f < 0$ has been significantly reduced (the tone becomes bluish rather than greenish). Bottom panel: similar to the top panel but including only low mass galaxies ($M_\star<$\,$10^7\,{\rm M_\odot}$). As expected, the fraction of potentials with $f<0$ increases (the tone becomes reddish) due to the increase of fits with $c<$\,0. The color bar shows \% of galaxies with  $f \geq 0$.
}
\label{fig:eddington2_plot_d}
\end{figure}
\subsection{Variable inner slope fits}\label{sec:intermediate}
Going a step further, we no longer force a core in the stellar density profile and the fits allows for a inner slope different from zero. The shape of the stellar density profiles is taken from Eq.~(\ref{eq:rhoc}) which scans from cored to NFW profiles when the inner slope $c$ goes from 0 to 1.
Examples of these fits are shown as green dashed lines in Fig.~\ref{fig:fitgood_2bc}, which contains exactly the same galaxies as Fig.~\ref{fig:fitgood}. In general, the fits using projected polytropes seem to do a slightly better job than these other fits, even though the number of free parameters is the same in both cases: $m$ has been replaced with $c$.   
This may be due to the fact that polytropes, rather than $\rho_{abc}$ profiles, are to be expected when self-gravitating systems reach thermodynamic equilibrium \citep{1993PhLA..174..384P}.  
Figure~\ref{fig:eddington3_plot} shows a scatter plot between the Relative RMS obtained with the two types of fit. Even if the differences are not overwhelming, the polytropic fits tend to be better than for these other fits. The fraction of good fits (Relative~RMS $<$ 0.02) is larger for polytropes than for the variable inner slope fits (44\,\% versus 32\,\%). Considering only good and fair fits, polytropic fits have lower Relative RMS in 59\,\% of the cases. Finally, the number of times where the polytropic fits are significanly better than the alternative
(defined as ratio of Relative RMS $>$ 1.5) is much larger than the oposite (19 galaxies versus 3 galaxies).

The properties of the variable inner slope fits are summarized in Figs.~\ref{fig:eddington2_plot_new_b} and \ref{fig:eddington2_plot_new_c}. The top panel of  Fig.~\ref{fig:eddington2_plot_new_b} shows the dependence of the inner slope $c$ as a function the Relative RMS. Note that often the best fitting $c$ is negative. Most of these negative slopes may be produced by systematic errors in the fit, a conjecture supported by two facts: (1)  the number of galaxies with $c<0$  increases with increasing Relative RMS (Fig.~\ref{fig:eddington2_plot_new_b}, top panel) and (2)  polytropic fits, where $c=0$ is imposed, often provide better fits, as discussed in the previous paragraph. The middle  panel of  Fig.~\ref{fig:eddington2_plot_new_b} contains the size of the core versus the goodness of the fit. No obvious trend is found. The bottom  panel of  Fig.~\ref{fig:eddington2_plot_new_b} shows the stellar mass versus the goodness of the fit indicating a hint of trend where more massive galaxies show better fits. The top panel of  Fig.~\ref{fig:eddington2_plot_new_c} shows a scatter plot of the inner slope $c$ as a function the stellar mass.  There is no clear trend.  The middle panel of Fig.~\ref{fig:eddington2_plot_new_c} displays the variation of the core radius with stellar mass.  More massive galaxies tend to have larger cores, as  it happened with polytropic fits (Fig.~\ref{fig:eddington1_plot_new_c}, middle panel). The bottom panel of Fig.~\ref{fig:eddington2_plot_new_c} shows the variation of the absolute RMS with stellar mass.

The diagnostic diagrams to constrain the properties of the gravitational potential consistent with these fits are shown in Fig.~\ref{fig:eddington2_plot_d}. Contrary to what happens with the polytropic fits, where most of the potentials are either allowed or forbidden (Fig.~\ref{fig:eddington1_plot_d}), most of the potentials are sometimes allowed and sometimes forbidden.
The results for all galaxies are shown in the top panel of Fig.~\ref{fig:eddington2_plot_d}. Around 50\,\% of the  potentials are not allowed, with a formal error in this estimate of the order of 5\,--\,10\,\% (Appendix~\ref{app:errors}). This happens irrespectively of the inner slope of the potential $c_p$, with the fraction increasing with increasing $r_s/r_{sp}$. As we discussed above  (Fig.~\ref{fig:eddington2_plot_new_b}, top panel), a significant part of the fitted profiles have $c \leq  0$, and these profiles are inconsistent with any potential. The  middle panel of Fig.~\ref{fig:eddington2_plot_d} includes only good fits (Relative RMS $<0.02$). The fraction of unphysical potentials is significantly reduced as expected following the reduction of objects with $c<0$ (Fig.~\ref{fig:eddington2_plot_new_b}, top panel). The bottom panel of Fig.~\ref{fig:eddington2_plot_d}  shows only low mass galaxies ($M_\star<$\,$10^7\,{\rm M_\odot}$). This other time the fraction of unphysical potentials increases due to the increase of the percentage of fits with $c<$\,0 among low-mass objects. 
Thus, we suspect the diagnostic plots to be significantly biased by the large uncertainty in $c$, which either allows for many potentials to be consistent ($c$ significantly larger than 0) or inconsistent ($c < 0$). The shape chosen for fitting does not constrain well enough the central slope of the observed galaxy. Or equivalently, the errors (stochastic and systematic) are still not small enough to reach a solid conclusion. We note that the scenario does not change significantly even when the Relative RMS is chosen as small as $<0.01$ (Fig.~\ref{fig:eddington5_plot_a}).
In the next section, we try another type of fit to check whether it improves the situation. The fitting functions used here tie the inner and outer slopes of the density profile, and this artificial bound may be causing significant systematic errors.

%
%
%
\begin{figure}
\centering
\includegraphics[width=0.9\linewidth]{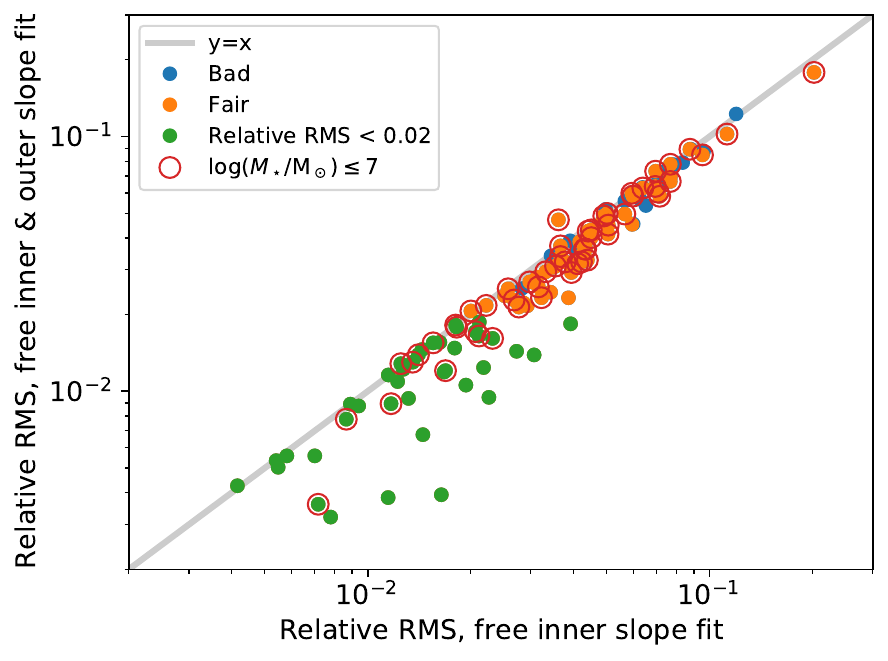}
\includegraphics[width=0.9\linewidth]{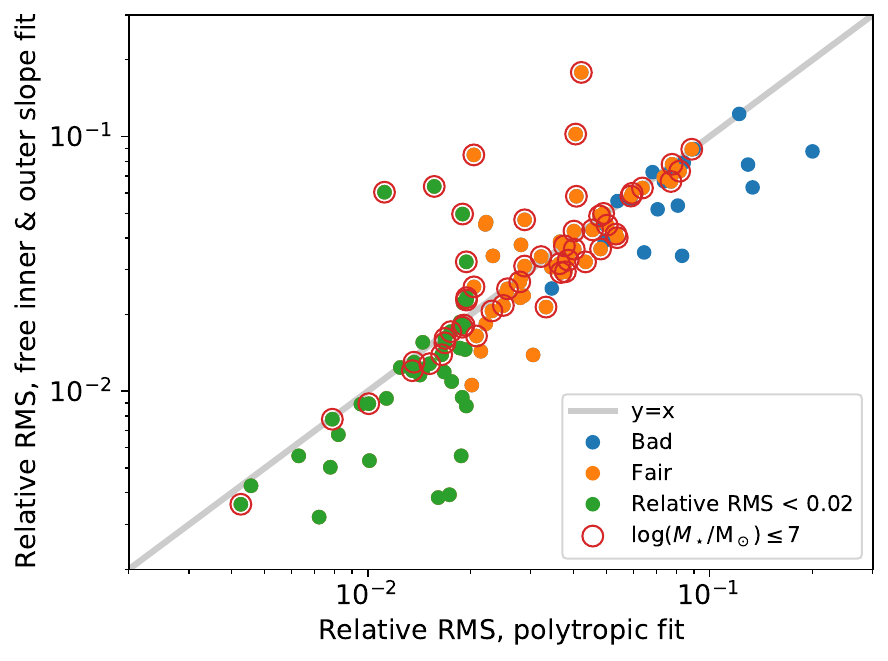}
\caption{Top panel: Scatter plot Relative RMS from the variable inner and outer slope fits versus the variable inner slope fits.  It is clear that the variable inner and outer slope fits generally has lower RMS, as expected since the underlying shape is the same and the former has one more degree of freedom.
Bottom panel:  Scatter plot Relative RMS from the variable inner and outer slope fits versus polytropic fits.
The colors and symbols are the same as in  Fig.~\ref{fig:eddington3_plot}.
}
  \label{fig:eddington3_plot_d}
\end{figure}
\begin{figure}
\includegraphics[width=\linewidth]{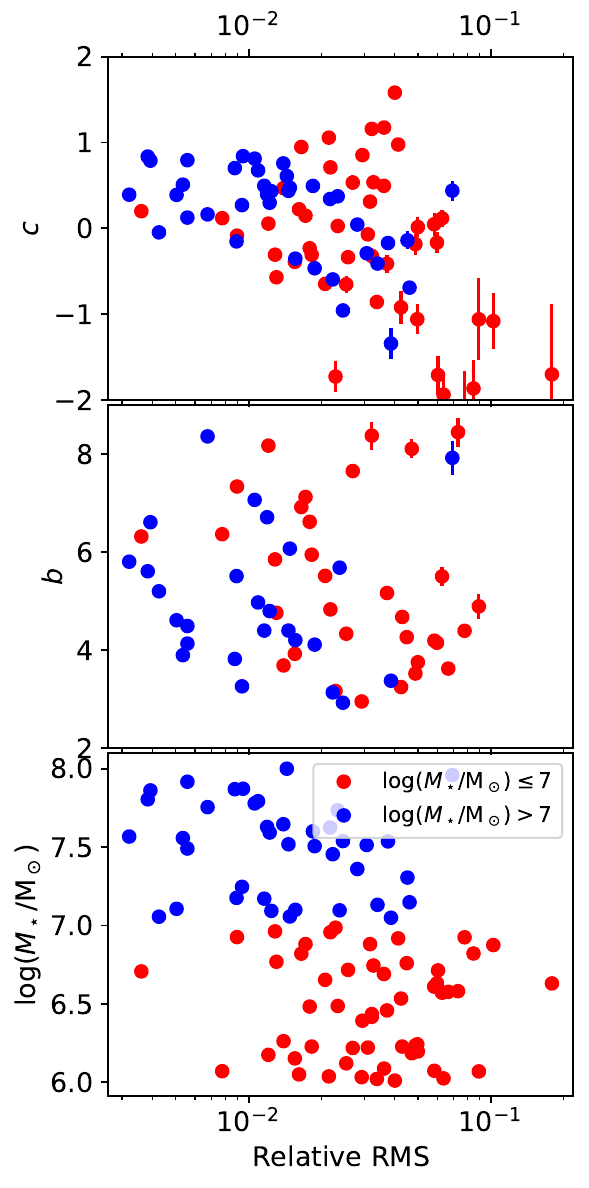}
\caption{Summary of the results with free inner ($c$) and outer ($b$) slopes presented as a function of the Relative RMS of the fit. To be compared with Fig.~\ref{fig:eddington1_plot_new_b}, for the polytropes, and Fig.~\ref{fig:eddington2_plot_new_b}, for fits when only the inner slope is free. The color code separates low and high stellar mass galaxies, as indicated in the inset.
Top panel: Inner slope. See how often $c < 0$, particularly for poor fits. 
Middle panel: Outer slope. For reference, $b=5$ in the case of polytropic fits with $m=5$, which is the limit $m$ for good polytropic fits (Fig.~\ref{fig:eddington1_plot_new_b}).
Bottom panel: Stellar mass. Lower mass galaxies tend to have  poorer fits.}
  \label{fig:eddington3_plot_new_b}
\end{figure}
\subsection{Variable inner and outer slopes}\label{sec:full}
Even though the variable inner slope profiles have the same number of free parameters as the polytropic fits, they tend to provide slightly worst fits (Fig.~\ref{fig:eddington3_plot}). The problem could be due to the fact that the inner slope in these fits, $c$, also sets the outer slope, $b=5-2c$  (c.f. Eqs.~[\ref{eq:rhoabc}] and [\ref{eq:rhoc}]). This tie minimizes the number of free parameters in the fit, but there is no clear physical reason to expect it.  To overcome the potential problem created by the artificial link between $b$ and $c$, we carried out additional fits where both the inner and outer slopes in Eq.~(\ref{eq:rhoabc}) are free parameters. The third parameter of the $abc$ profiles, $a$, which determines the transition between the inner and outer parts, are set to 2, which is the value corresponding to the polytropes\footnote{We also tried $a=1$ but the fits are worse than those for $a=2$ reported here.} (Eq.~[\ref{eq:polapprox}]). Examples of fits with free inner and outer slopes are shown as orange lines in Fig.~\ref{fig:fitgood_2bc}. The figure also includes the fits with variable inner slope only (green dashed lines). Even though the Relative RMS is lower when both $c$ and $b$ are free  parameters (top panel, Fig.~\ref{fig:eddington3_plot_d}), the fitted profiles look very similar, and also similar to the observed profiles   (Fig.~\ref{fig:fitgood_2bc}). Figure~\ref{fig:eddington3_plot_new_b} shows how $c$ and $b$ depend on the Relative RMS of the fit.

%
%
The constraints on the gravitational potential imposed by these fits are similar to those when only the inner slope is free (Fig.~\ref{fig:eddington2_plot_d}). There is no major qualitative difference. A significant part of the fits still render $c \leq 0$ and therefore unphysical $f < 0$ DFs. The galaxies can be divided into those with $c \gtrsim 0.1$, which are consistent with a NFW potential ($c_p=1$) and those with  $c \lesssim  0.1$ which are inconsistent (see Sect.~\ref{sec:eddington_summary}). However, part of the scatter in $c$ is probably artificial as deduced from the fact that the scatter increases with increasing Relative RMS (Fig.~\ref{fig:eddington3_plot_new_b}, top panel). The uncertainties in the fits are partly due to errors in the photometry but, most likely, the are systematic arising from deviations from the assumptions implicit in the fits, including spherical symmetry, $\rho_{abc}$ profile, and $a=2$.
We have tried to estimate the impact of these systematic errors on the inner slope $c$ under the working hypothesis that all negative slopes  $c<0$ are due to noise. We select all those profiles with good fits (Relative RMS $<0.02$) and $c < 0$. If these values are random excursions produced by noise of profiles with $c \simeq 0$, then the standard deviation of the noise would be $\sigma_c \simeq 0.3$. (The same exercise with the fits in Sect.~\ref{sec:intermediate}, where only the inner slope was fitted, renders $\sigma_c\geq 0.25$.) Thus, we find that 38\,\% of the good fits have $|c| \leq\sigma_c$ and therefore are consistent with $c=0$ within errors. (The same exercise with the inner slope free fits renders 60\,\%.) In other words,  around 40\,\% of the fits are consistent with cores in the stellar mass distribution (i.e., with $c=0$) and so likely inconsistent with NFW-like potentials. This figure is also similar to the limit steaming from the diagnostic plot (e.g., Fig.~\ref{fig:eddington2_plot_d}, middle panel).  

%
The  Relative RMS of the free inner and outer slope fits is in general smaller than for polytropes (Fig.~\ref{fig:eddington3_plot_d}, bottom panel). However, the comparison of the two types of fits is not direct since two-slopes fits have one more degree of freedom than polytropes, and since the two shapes are different, an increase of the number of free parameters my artificially reduce the RMS because of overfitting. This being said, we find that the number of profiles with  Relative RMS $< 0.02$ is very similar in both cases (41 for polytropic fits  versus 39 for $\rho_{2bc}$) so is the number of profiles with significantly better fit of one relative to the other (10 versus 12). The similitude between the numbers characterizing both types of fit supports the above conclusion that the observed profiles are often consistent with $c=0$. The core ($c=0$) is imposed for polytropes but inferred for fits with free inner and outer slopes.

%
\section{Conclusions}\label{sec:conclusions}

\citet{2023ApJ...954..153S} proposed the use of the classical Eddington Inversion Method (EIM) to constrain the properties of the gravitational potential in galaxies using only the observed distribution of starlight. As we explain in Sect.~\ref{sec:intro}, the procedure may be able to
  decipher the inner shape of the DM halos
in the ultra-low mass regime where (1) the potential is set by the DM, and (2) the baryon feedback processes are unable to transform cuspy DM halos (typical of collisionless DM) into cored halos. Thus, in this mass regime, the finding of cores in the DM distribution would imply that the DM is not collisionless. Even though dynamical measurements of the DM distribution in these galaxies are unfeasible, the EIM may allow them, thus opening a new possibility to explore the nature of DM.

The algorithm employed in the paper is detailed in Sect.~\ref{sec:procedure} and tested using GCs in Appendix~\ref{app:gctests}.
Given an observed stellar mass distribution, we check whether or not it is consistent with a battery of gravitational potentials. The inconsistency manifest itself when, for the stellar mass to reside in a particular potential, the distribution function in the phase space $f$ has to adopt negative values in some regions of the phase space. 
The EIM allows us to compute $f$ under the assumption that the system is spherically symmetric with isotropic velocities. \modified{Since these assumptions seem quite restrictive, we discuss some of the limitations they impose in Sect.~\ref{sec:procedure} and Appendix \ref{app:extension}.}
Here we present the first practical application of the technique using surface brightness profiles of around 100 low mass galaxies ($10^6\,{\rm M_\odot} \leq M_\star\leq 10^8\,{\rm M_\odot}$) from \citet{2021ApJ...922..267C, 2022ApJ...933...47C}, who selected them as dwarf satellites of Milky Way-like galaxies (see Sect.~\ref{sec:observations}). Their masses are still too large to constrain the nature of DM but the set provides an excellent testbed for the technique in terms of number of targets and noise level.   
The technique depends on the type of function used for fitting the observed profiles. We try three types: polytropes, profiles with free inner slope (Eq.~[\ref{eq:rhoc}]), and profiles with free inner and outer slopes (Eq.~[\ref{eq:rhoabc}], with $a=2$). The application of the technique allows us to reach the following conclusions:

\begin{itemize}
\item The observed profiles are generally well fitted by projected polytropes, with 41 galaxies ($44$\,\% of the total) having pretty low residuals with a Relative RMS  (Eq.~[\ref{eq:relative_rms}]) $< 0.02$. The polytropic index tends to 5 for the best fits and the derived core radius scales with galaxy stellar mass (Fig.~\ref{fig:eddington1_plot_new_b}). The good fits have a RMS $\sim 0.03$\,dex, equivalent to 7\,\% in flux  (Fig.~\ref{fig:eddington1_plot_new_c}).
\item As expected (Sect.~\ref{sec:eddington_summary}), NFW-like potentials are discarded if the polytropic shape is good enough to reproduce the inner parts of the galaxies (Fig.~\ref{fig:eddington1_plot_d}).
\item More massive (more luminous) galaxies tend to have better fits, supporting that noise (random and systematic) is affecting the results. The trend is present for the three types of fits (Figs.~\ref{fig:eddington1_plot_new_b}, \ref{fig:eddington2_plot_new_b}, and \ref{fig:eddington3_plot_new_b}, bottom panels).
\item When a parameterized profile with variable inner slope is used for fitting, then the quality of the fits worsens with respect to the polytropes, even though both profiles have the same number of free parameters (Fig.~\ref{fig:eddington3_plot}). In this case, the inner slope $c$ is often negative (Fig.~\ref{fig:eddington2_plot_new_b}), and the diagnostic diagrams basically discard all potentials if $c\lesssim 0.1$ (Fig.~\ref{fig:eddington2_plot_d}).
\item When that inner and outer slopes are allowed to vary simultaneously (Sect.~\ref{sec:full}), the quality of the fits improves (Fig.~\ref{fig:eddington3_plot_d}, left panel), but the fact that $c$ is often negative remains (Fig.~\ref{fig:eddington3_plot_new_b}, top panel). 
Some of theses negative slopes may be produced by systematic errors in the fit, a conjecture supported by the fact that the number of galaxies with $c < 0$ increases with increasing Relative RMS (Fig.~\ref{fig:eddington3_plot_new_b}, top panel). 
Assuming that the negative excursions of $c$ are due to errors in the fit, we estimate the error in $c$  to be $\sigma_c\simeq 0.3$, which implies that around 40\,\% of the good fits are consistent with $c=0$ (Sect.~\ref{sec:full}).
\item If polytropes were the correct fits (which is likely the case; see Sect.~\ref{sec:full}) then all galaxies are inconsistent with NFW-like potentials (Fig.~\ref{fig:eddington1_plot_d}). In the case of allowing the inner slope to vary (Sects.~ \ref{sec:intermediate} and \ref{sec:full}), between 40\,\% and 70\,\% of the galaxies are consistent with cores in the stellar mass distribution ($c=0$) thus inconsistent with NFW-like potentials (Fig.~\ref{fig:eddington2_plot_d}).
\item  The EIM tools that we developed and use assume isotropic velocities. However, the inner slope of a stellar core and the radial anisotropy $\beta$ (Eq.~[\ref{eq:ani-param}]) are related so that $c > 2\beta$ for the system to be physically realizable.  In other words, large radially biased orbits ($\beta > 0$) are strongly inconsistent with soft stellar cores even if $c$ is significantly larger than zero. In low mass galaxies, $\beta > 0$ is to be expected  \citep[e.g.,][]{2017ApJ...835..193E,2021MNRAS.504.3509O} and so the limits on $c$ in the previous item are likely to be conservative.  
\end{itemize}

The above results have been derived under the assumption that the observed stellar systems are spherically symmetric. However, dwarfs tend to be oblate or even triaxials \citep[e.g.,][]{2016ApJ...820...69S,2019ApJ...883...10P}. The construction of the 1D profiles in Fig.~\ref{fig:raw_prof} includes symmetrizing the observed profiles, but it is unclear to what extent this artifice is enough to grant the application of an approach which,  strictly speaking, is valid only for spherically symmetric systems. Fortunately, this assumption does not seem to be critical since it does not compromise the main observational constraint leading to the results, namely, the existence of stellar cores.  Firstly, all good observed profiles converge to the same shape, a polytrope of index $m\sim 5$ (Sect.~\ref{sec:pp_results}). Thus, since this shape seems to be independent of the original axial ratio,  it is also the one expected in purely spherical stellar system, for which our analysis applies. Secondly, one can show (Appendix~\ref{app:extension}) how the incompatibility between NFW potentials and stellar cores also remains for axi-symmetric systems  (e.g., oblate), therefore, it is more fundamental than, and not attached to, the spherical symmetry assumption.

In short, the EIM-based technique seems to work in the sense of providing constraints on the gravitational potential using only good but typical observed surface brightness profiles. Our tests also indicate that good data are needed to break the degeneracies in the determination of the inner slope of the profiles. However, since the key parameter is the inner slope,  reaching the innermost regions is more important than decreasing noise. Thus, observing with enough spatial resolution is critical. Having large samples of objects is also important, so that those deviating from the EIM assumptions can be discarded without compromising the large statistics needed to infer general conclusions on the nature of DM.

Even if this is not mentioned explicitly, the gravitational potential to be used in the EIM also includes the contribution from baryons. All the above conclusions still hold provided the total density of the system (i.e., stars, gas, and DM included) is well approximated by the density profile used to construct the battery of potentials. Equation~(\ref{eq:rhoabc}) was used in this paper, but this is not a serious restriction of the technique since other shapes can be incorporated if needed.


%

\begin{acknowledgements}
  Thanks are due to Scott Carlsten who generously provided the surface density profiles analyzed in the paper and to Barbara Lanzoni and Francesco Ferraro for the globular cluster profiles used in Appendix~\ref{app:gctests}.
Thanks are also due to an anonymous referee who suggested clarifications leading to the extension to non-spherically symmetric systems in Appendix~\ref{app:extension} and the sanity check in Appendix~\ref{app:gctests}.
  JSA acknowledges financial support from the Spanish Ministry of Science and Innovation, project PID2022-136598NB-C31 (ESTALLIDOS), and from Gobierno de Canarias through EU FEDER funding, project PID2020010050. His visit to La Plata was partly covered by the MICINN through the Spanish State Research Agency, under Severo Ochoa Centers of Excellence Programme 2020-2023 (CEX2019- 000920-S).
IT acknowledges support from the ACIISI, Consejer\'{i}a de Econom\'{i}a, Conocimiento y Empleo del Gobierno de Canarias and the European Regional Development Fund (ERDF) under grant with reference PROID2021010044 and from the State Research Agency (AEI-MCINN) of the Spanish Ministry of Science and Innovation under the grant PID2022-140869NB-I00 and IAC project P/302302, financed by the Ministry of Science and Innovation, through the State Budget and by the Canary Islands Department of Economy, Knowledge and Employment, through the Regional Budget of the Autonomous Community.
We acknowledge the use of the Python packages {\em numpy} \citep{2020Natur.585..357H}, {\em scipy} \citep{2020NatMe..17..261V}, and {\em matplotlib} \citep{Hunter:2007}.

\end{acknowledgements}

%

   \appendix
   \section{Error bars for the diagnostic plots}\label{app:errors}

\begin{figure}
  \centering
\includegraphics[width=0.9\linewidth]{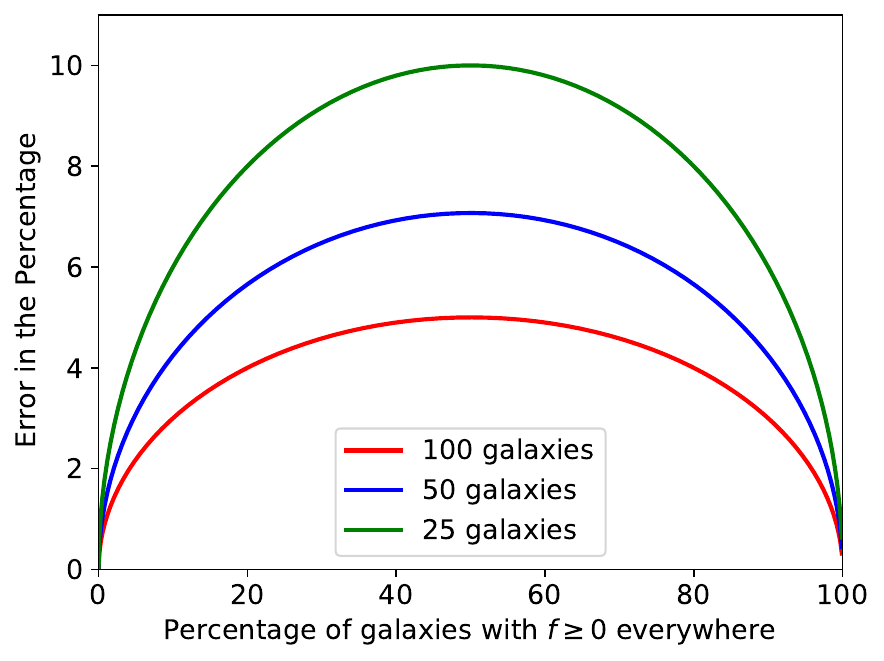}
\caption{Errors expected in the diagnostic diagrams showing percentage of galaxies with $f\geq 0$ everywhere. It depends on the number of galaxies used for diagnostics, as indicated in the inset.  
  }
  \label{fig:appa}
\end{figure}

   The diagnostic plot introduced in Sect.~\ref{sec:diagrams} and shown in Figs.~\ref{fig:eddington1_plot_d} and \ref{fig:eddington2_plot_d} is based on counting the number of galaxies that, for the same potential, do or do not have a particular property (i.e., whether or not $f \geq 0$ everywhere). Thus, the number of counts $n$ should follow a binomial distribution characterized by the probability $q$ of having the property \citep[e.g.,][]{1971stph.book.....M}. In this case, the expected value of counts is $qN_g$, with $N_g$ the number of galaxies. The error in the number of counts, computed as the square root of the variance of the distribution, is  $\sqrt{q(1-q)N_g}$. Thus, the percentage of galaxies represented in the diagnostic plot, $g= 100\times n/N_g$, has an error $\sigma_g$ of
\begin{equation}
\sigma_g = 100\times\sqrt{\frac{q(1-q)}{N_g}}.
\end{equation}
The error in the percentage versus the measured percentage is shown in Fig.~\ref{fig:appa} for three representative values of $N_g$ (25, 50, and 100). Errors are largest when the percentage is 50\,\% , and tend to zero at 0 and 100\,\%.  Their values for the characteristic $N_g\sim 50$ are typically $\lesssim$\,6\,\%.

\section{Incompatibility between stellar cores and NFW potentials beyond spherically symmetric systems}\label{app:extension}

This appendix proves that the incompatibility between NFW potentials and stellar cores goes beyond the assumption of spherical symmetry, being more fundamental than and not attached to this particular hypothesis.

  \citet{1962MNRAS.123..447L} extended the EIM to axi-symmetric systems. The axis of symmetry of the stars is assumed to be the same as the potential. In this case, the even component of the distribution function, $f_+$, depends on two integrals of motion, the relative energy $\epsilon$ and the angular momentum along the axis of symmetry. The density depends on the cylindrical coordinates $R$ (radius) and $z$ (distance along the axis of symmetry). Using the potential $\Psi$ to parameterize the dependence on $z$ at a fixed $R,$ one can write down an equation \citep[see][Sect.~4.4.1]{2008gady.book.....B}, 
\begin{equation}
\frac{\partial\rho(R,\Psi)}{\partial\Psi}=2\pi\sqrt{2}\int_0^{\Psi}\frac{f_+\left[\epsilon,2(\Psi-\epsilon)R^2\right]}{\sqrt{\Psi-\epsilon}}d\epsilon, 
\end{equation}
which is formally identical to Eq.~(\ref{derodepsi}). This equation, particularized to $R=0$, can be used to compute the derivative of the density along $z$, rendering, 
\begin{equation}
\frac{\partial\rho(0,\Psi)/\partial z}{\partial\Psi/\partial z}=2\pi\sqrt{2}\int_0^{\Psi}\frac{f_+\left[\epsilon,0\right]}{\sqrt{\Psi-\epsilon}}d\epsilon.
\end{equation}
As it happens with $f$, $f_+$ has to be non-negative everywhere in the phase space \citep[][Eq.~4.39b]{2008gady.book.....B}, therefore, the arguments given in Sect.~\ref{sec:eddington_summary} apply here as well. If there is a stellar core  
along the $z$ direction, i.e.,
\begin{equation}
  \partial\rho(0,\Psi)/\partial z=0 {\rm~when~} z\to 0,
\end{equation}
then for $f_+ \geq 0$  everywhere,
\begin{equation}
\partial\Psi/\partial z=0 {\rm~when~} z\to 0,
\end{equation}
which is incompatible with $\Psi$ being a NFW potential and therefore stellar cores and NFW potential are physically incompatible also in axi-symmetric systems.
As we stress above, this fact proves that the incompatibility between NFW potentials and stellar cores does not arise from the spherical symmetry assumption.

\section{Sanity check based on Globular Clusters}\label{app:gctests}
\begin{figure}
\centering
\includegraphics[width=0.95\linewidth]{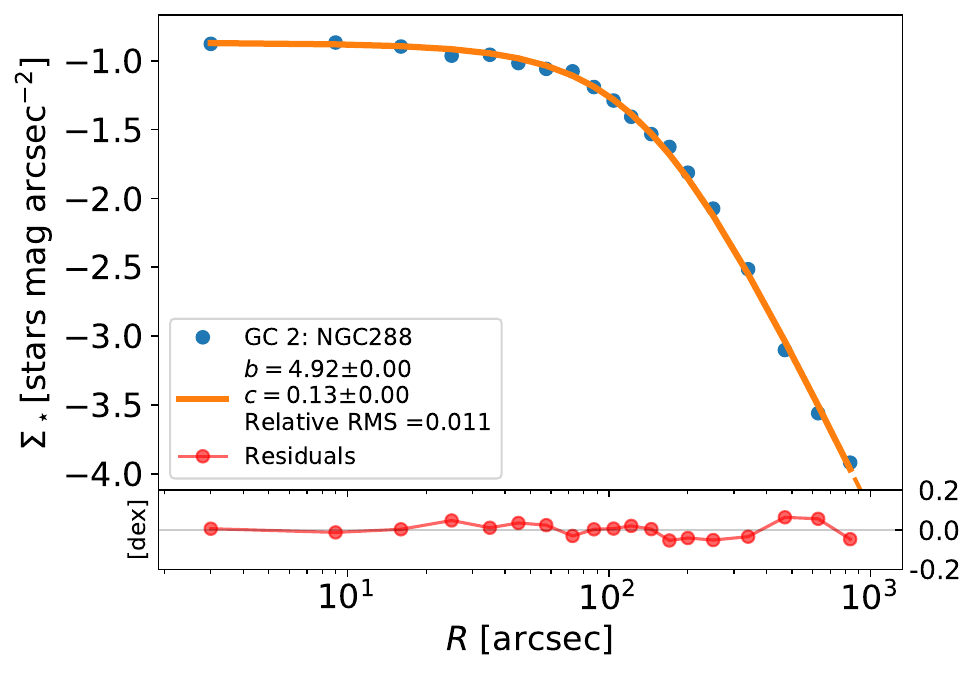}
\includegraphics[width=0.95\linewidth]{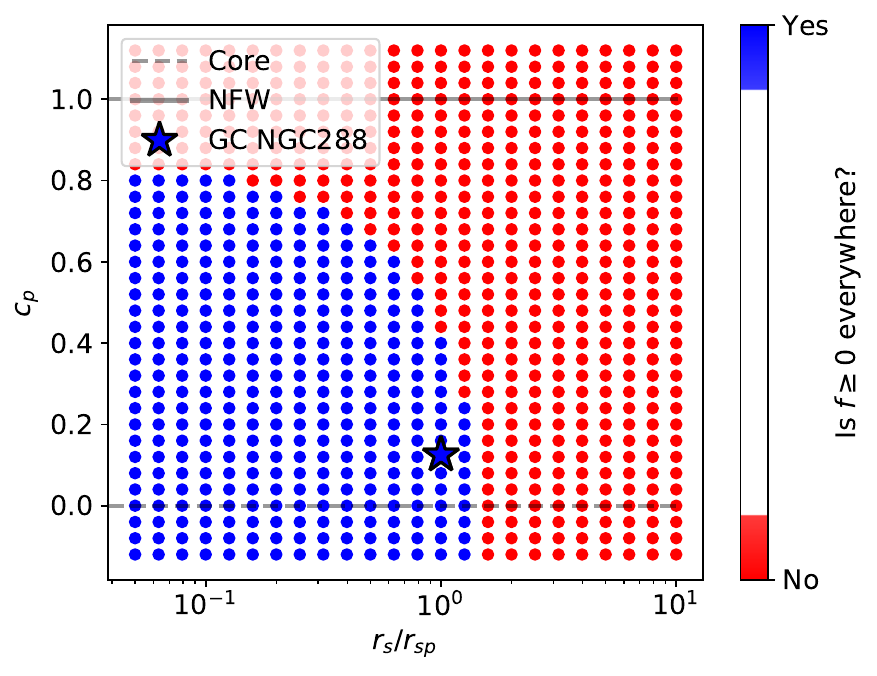}
\includegraphics[width=0.95\linewidth]{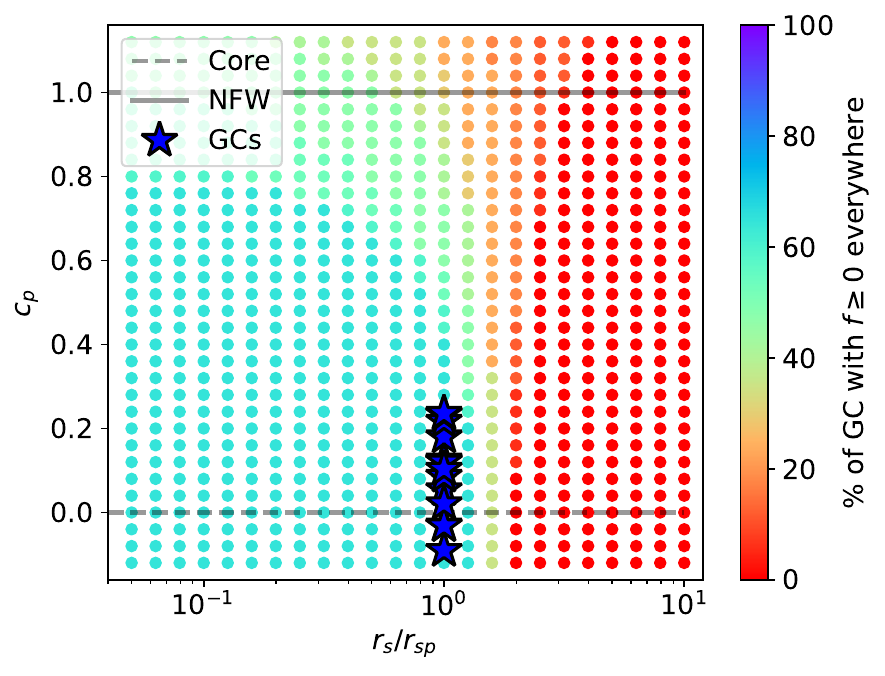}
\caption{
Sanity check employing globular clusters \citep[from][]{2013ApJ...774..151M}, thought to be self-gravitating systems. Top panel: fit to one of them, NGC\,288. Middle panel: diagnostic diagram for NGC\,288 inferred from the above fit.  Because it is self-gravitating, the parameters of the stellar distribution (indicated by the star symbol) are those of the true gravitational potential ($r_{sp}$ and $c_p$), and they appear in the allowed region of the diagnostic diagram. Bottom panel:  diagnostic diagrams combining all GCs together. As in the middle panel, the star symbols indicated the true potentials, and they all fall well within the allowed region in the diagnostic plot.
}
  \label{fig:eddington13_plot_d}
\end{figure}

    Globular clusters (GC) are known to be self-gravitating stellar systems, with little or no DM \citep[e.g.,][]{2012ApJ...755..156S,2013MNRAS.428.3648I}, therefore, we know their gravitational potential directly from the observed stellar distribution. Moreover, GCs usually show inner cores produced by the thermalization of the system through two-body gravitational interactions between stars \citep[e.g.,][]{2008gady.book.....B}. Thus, GCs provide excellent testbeds for the EIM-based approach followed in the paper. This appendix presents one such test using the GCs observed with the Hubble Space Telescope  by \citet[][]{2013ApJ...774..151M}, with their mass surface density  profiles derived from resolved stellar counts. The tools introduced in Sect.~\ref{sec:procedure} are applied to them. Specifically, the fits allow for free inner and outer slopes, as described in Sect.~\ref{sec:full}. From the original 26 GCs, we keep 17 having good fits to the innermost radii with cores. An example is shown in the top panel of Fig.~\ref{fig:eddington13_plot_d}, which is equivalent to any panels in Figs.~\ref{fig:fitgood} or \ref{fig:fitgood_2bc}. The corresponding diagnostic diagram (as in Fig.~\ref{fig:eddington1_plot_d0}) is shown in the middle panel of Fig~\ref{fig:eddington13_plot_d}, where the large star symbol shows the inner slope and characteristic radius of the GC (upper panel in Fig.~\ref{fig:eddington13_plot_d}). If the GC  is self-gravitating, these two parameters from the stars define the gravitational potential. One can see how the parameters from the fit appear in the allowed region of the diagnostic diagram (indicated by the blue bullets). Moreover, the diagnostic plot completely discards NGC\,288 residing in a NFW potential (i.e., discards $c_p=1$ for $r_s=r_{sp}$).
Repeating the exercise with the other GCs renders the lower panel in  Fig.~\ref{fig:eddington13_plot_d}, which is equivalent to Figs.~\ref{fig:eddington1_plot_d0} and \ref{fig:eddington1_plot_d} in the main text.  Again, the star symbols representing the true potentials fall well within the allowed region in the diagnostic plot, and have harder times to be consistent with NFW potentials.    
Thus, the tools employed in the paper give consistent results when applied to GCs, thought to be self-gravitating structures and therefore having known gravitational potentials.

\end{document}